\documentclass[11pt,english]{article}
\usepackage[T1]{fontenc}
\usepackage[latin9]{inputenc}
\usepackage{geometry}
\usepackage{amsmath,amsfonts,epsfig}
\usepackage{amssymb}
\usepackage{jheppub}
\usepackage{mathrsfs}
\usepackage{hyperref}
\usepackage{hyperref}
\usepackage{comment}
\newcommand{\CP}{\mathop{\textrm{CP}}}
\newcommand{\mm}{{\mathfrak{m}}}
\newcommand{\scL}{{\cal L}}
\newcommand{\scG}{{\cal G}}
\newcommand{\al}{\alpha}

\newcommand{\tom}{\tilde{\omega}}
\newcommand{\om}{{\omega}}
\newcommand{\de}{{\delta}}
\newcommand{\ltl}{{\log\tfrac{2}{\ell}}}

\newcommand{\intinf}{\mathop{\int\limits_{-\infty}^\infty}}

\newcommand{\intli}{\mathop{\int\limits_{-\log\tfrac{2}{\ell}}^\infty}}

\newcommand{\intil}{\mathop{\int\limits_{-\infty}^{\log\tfrac{2}{\ell}}}}

\begin{document}
\numberwithin{equation}{section}
\setlength{\unitlength}{.8mm}



\title{Leading UV Formula for Finite-Volume Vertex Operator Expectation Values in the Sine-Gordon Model from Kink NLIE
}

\author[a]{\'Arp\'ad Heged\H us,}
\author[a,b]{Apor Roth}

\affiliation[a]{
HUN-REN Wigner Research Centre for Physics,\\
H-1525 Budapest 114, P.O.B. 49, Hungary}

\affiliation[b]{
Roland E\"otv\"os University, Department for Theoretical Physics, \\
1117 Budapest, P\'aym\'any P\'et\'er s\'et\'any 1/A, Hungary
}

\emailAdd{hegedus.arpad@wigner.hun-ren.hu}
\emailAdd{apor.roth@wigner.hun-ren.hu}

\abstract{

We study the ultraviolet (UV) limit of finite-volume expectation values of vertex operators in the sine-Gordon model using the kink nonlinear integral equation (NLIE) description of the conformal limit.

By analysing the integrable formulation of vacuum expectation values in the small-volume regime, we conjecture an explicit analytic expression for the leading asymptotic term in the small-volume expansion, formulated in terms of kink functions. This establishes a direct connection between the integrable finite-volume description and the expected conformal asymptotics determined by the 3-point functions of the underlying conformal field theory (CFT).

The proposed formula is tested against the analytic expression known from complex Liouville conformal field theory using high-precision numerics, showing agreement to at least 19 significant digits.

}

\maketitle

\section{Introduction } \label{intro}


Solving conformal field theories (CFTs) remains a central challenge in many branches of theoretical physics, from string theory to statistical mechanics. Conformal symmetry imposes strong constraints, and once the fundamental CFT data - scaling dimensions and 3-point structure constants - are known, all correlation functions in principle become accessible.

Determining this data, however, is in general highly nontrivial and in most cases remains out of reach. 
In two dimensions, the infinite-dimensional Virasoro algebra \cite{Belavin:1984vu} provides a powerful analytic framework for constructing exact CFT data. 
In higher-dimensional theories, progress relies predominantly on the conformal bootstrap programme, both in its analytic \cite{Poland:2022qrs} and numerical \cite{Poland:2018epd} implementations, which has led to precise constraints on universal CFT data.

A particularly remarkable exception is provided by planar theories arising in AdS/CFT dualities \cite{Maldacena:1997re}, where integrability plays a central role \cite{Beisert:2010jr}. 
In the case of the AdS$_5$/CFT$_4$ correspondence, the spectral problem of $\mathcal{N}=4$ supersymmetric Yang--Mills theory has been essentially solved through integrability-based methods, most notably the Y-system formulation \cite{Gromov:2009tv}, and its modern reformulation in terms of the Quantum Spectral Curve (QSC) \cite{Gromov:2013pga,Gromov:2014caa}. 
On the other hand, while the computation of three-point functions has seen substantial progress within the hexagon framework \cite{Basso:2015zoa}, the inclusion of all wrapping and gluing effects remains challenging. 
For CFTs endowed with integrable structures, it is natural to expect that three-point couplings 
admit relatively compact representations 
in terms of functions characterizing the integrable description of the spectral problem of the underlying  system.

In the context of the AdS$_5$/CFT$_4$ spectral problem, techniques originating from two-dimensional integrable quantum field theory have provided important guiding principles for the development of methods used to determine the spectrum. 
This suggests that integrability-based approaches to three-point functions in two-dimensional CFTs may also provide structural insights relevant to the AdS/CFT correspondence.  


The aim of the present work is to develop a framework for representing three-point couplings in two-dimensional CFTs directly in terms of functions that emerge from the integrable description of the spectrum.

This programme was initiated in \cite{Bajnok:2022ucr}, where the Lee-Yang and three-state Potts models were studied.
The analysis was restricted to a particular class of three-point couplings involving two identical operators and a third operator corresponding to the perturbation that generates a massive integrable deformation of the CFT.
More precisely, couplings of the form $C_{{\cal O} \Phi {\cal O}}$ were considered, where ${\cal O}$ denotes the repeated operator and $\Phi$ the perturbing field.\footnote{The special case ${\cal O}=\Phi$ was also included.}

As a result, the method developed in \cite{Bajnok:2022ucr} provides access to only a limited subset of the full space of three-point couplings.
This approach determines these couplings by computing the UV limit of the expectation values $\langle {\cal O}|\Phi|{\cal O} \rangle$ in the massive perturbed theory.
The requirement that the sandwiched operator coincide with the perturbing field stems from the fact that these particular three-point couplings can be extracted from the ultraviolet (UV) behaviour of the finite-volume energy of the sandwiching state in the perturbed theory. 
Consequently, methods aimed solely at solving the spectral problem are sufficient to determine the corresponding CFT data.


In \cite{enjhep}, the analysis of \cite{Bajnok:2022ucr} was extended to a broader class 
of sandwiched operators (i.e., three-point couplings) in the ultraviolet limit of the sine-Gordon model. 
This extension was made possible by the framework developed in \cite{Jimbo:2010jv}, 
where, using the fermionic basis \cite{Boos:2007wv,Boos:2006mq,Boos:2008rh,Jimbo:2008kn,Boos:2010qii}, 
all finite-volume expectation values of the theory can, in principle, be computed from the integrable 
description of its spectral problem. While the formulas in \cite{Jimbo:2010jv} and \cite{Hegedus:2019rju} 
provide the volume dependence of the massive theory, obtaining the 3-point couplings in the 
limiting CFT requires taking the ultraviolet limit of these expressions in terms of integrability data.

A central result of \cite{Jimbo:2010jv} is that, within the integrable description, expectation values 
can be expressed in terms of a finite set of elements of an infinite matrix $\omega_{2k-1,1-2j}(\al)$, 
with $j,k \in \mathbb{Z}$.


In \cite{enjhep}, the first two nontrivial coefficients in the UV expansion of the $\omega$ matrix 
were computed, yielding an integrable formulation of the leading term in the UV expansion of 
a restricted class of ratios of vacuum expectation values, such as $\frac{\langle \Phi_{\al+2 \, \tfrac{1-\nu}{\nu}}\rangle}{\langle \Phi_{\al}\rangle}$ 
and $\langle \Phi_{4\tfrac{1-\nu}{\nu}}\rangle$, in the sine-Gordon model.{\footnote{For the definitions 
of these operators, see section \ref{2}.}} 
It was also emphasized in \cite{enjhep} that a complete determination of the leading UV behavior of all ratios of the form 
$\tfrac{\langle \Phi_{\al+2 m \tfrac{1-\nu}{\nu}}\rangle}{\langle \Phi_{\al}\rangle}$, for $m=1,2,\dots$, within the framework of \cite{Jimbo:2010jv}, 
would, at least naively, require an all-loop evaluation of the $\omega_{2k-1,1-2j}(\al)$ matrix.

In the present work, we complete this program by computing the leading term in the UV 
expansion of the general ratio $\tfrac{\langle \Phi_{\al+2 m \tfrac{1-\nu}{\nu}}\rangle}{\langle \Phi_{\al}\rangle}$. 
This is achieved by systematically identifying the relevant contributions at each order 
in the UV perturbative series of $\omega_{2k-1,1-2j}(\al)$. 
The resulting analysis yields a closed-form expression for the coefficient of the leading term in the UV expansion, valid for arbitrary values of the coupling constant and for all $m \in \mathbb{N}$. The result is expressed in terms of kink functions, which provide the integrable description of the model's ultraviolet limit.


The structure of the paper is as follows:

In Section \ref{2}, we review the connection between vacuum expectation values and 3-point couplings in a CFT, as well as key results on the integrable formulation of finite-volume expectation values of local operators in the sine-Gordon theory. Section \ref{3} is devoted to the analysis of the UV behavior of the solution to the nonlinear integral equation (NLIE) governing the finite-volume ground-state energy. In Section \ref{4}, we summarize the predictions for the UV behavior of expectation values derived from field-theoretical considerations. Section \ref{5} presents our final integrability-based expressions for the leading UV coefficient of the expectation value ratios $\tfrac{\langle \Phi_{\al+2 m \tfrac{1-\nu}{\nu}}\rangle}{\langle \Phi_{\al}\rangle}$ for $m=1,2,\dots$. The paper concludes with a brief discussion of the results in Section \ref{9}.

Technical details supporting our main results are collected in Appendix \ref{appA}, where we present the explicit analytic computations underlying our findings.
In Appendix \ref{appB}, we provide the large-argument series expansion of the principal integral kernel of the problem. Finally, Appendix \ref{appC} contains tables of high-precision numerical data, which serve to validate our closed - form expression for the leading term in the UV series of the vacuum expectation values of primary fields.


\section{Preliminaries } \label{2}

The central idea of our approach to investigating the conformal data of a CFT is to place the theory 
in finite volume, equivalently defining it on a cylinder of circumference $L$.  
In finite volume, the energy spectrum of a CFT is entirely determined by the scaling dimensions and the 
central charge of the model. Thanks to the state-operator correspondence, each energy level can be 
associated with a specific operator of the theory. The volume dependence of an energy level corresponding 
to an operator ${\cal O}$ is fully fixed by its conformal dimensions:
\begin{equation} \label{EcftO}
\begin{split}
E_{\cal O}(L)=\frac{2 \, \pi}{L} \left( \Delta_+^{\cal O} +\Delta_-^{\cal O} -\frac{c}{12} \right),
\end{split}
\end{equation}
where $\Delta_\pm^{\cal O}$ denote the left and right conformal dimensions of ${\cal O}$, and $c$ is 
the central charge of the CFT.

In finite volume, the matrix elements of local fields provide access to the 
3-point couplings of the theory via
\begin{equation} \label{cftexpL}
\begin{split}
\langle {\cal O}_1 | {\cal O}  | {\cal O}_2 \rangle=\left( \frac{2 \, \pi}{L} \right)^{\Delta_+^{\cal O} +\Delta_-^{\cal O}} 
C_{{\cal O}_1 {\cal O} {\cal O}_2}.
\end{split}
\end{equation}

In this work, we aim to develop an integrability-based description of 
the three-point couplings $C_{{\cal O}_1 {\cal O} {\cal O}_2}$ in the ultraviolet limit of the sine-Gordon theory. 
Due to limitations of the available analytical tools, we restrict ourselves to three-point couplings 
involving at least two identical operators (${\cal O}_1={\cal O}_2$ in (\ref{cftexpL})), 
which correspond to diagonal matrix elements in the finite-volume setting.

\subsection{The model}

In this work, we study the sine-Gordon model, defined by its Euclidean action as
\begin{equation} \label{LSG}
\begin{split}
{\cal A}_{SG}=\int \bigg\{ \frac{1}{4 \pi} \partial_z \varphi(z,\bar{z}) \, \partial_{\bar{z}} \varphi(z,\bar{z})
-\frac{2 {\bf \mu}^2}{\sin \pi \beta^2} \cos(\beta \, \varphi(z,\bar{z})) \bigg\} \, \frac{i \, dz \wedge d\bar{z}}{2},
\end{split}
\end{equation}
where $z=x+i\, y$ and $\bar{z}=x-i \, y$ denote the Euclidean space-time coordinates.

Following \cite{Jimbo:2010jv}, we treat the model as a perturbed complex Liouville CFT:
\begin{equation} \label{PCLT}
\begin{split}
{\cal A}_{SG}={\cal A}_{L}- \frac{{\bf \mu}^2}{\sin \pi \beta^2} \int e^{i \, \beta \, \varphi(z,\bar{z})} \frac{i \, dz \wedge d\bar{z}}{2},
\end{split}
\end{equation}
where ${\cal A}_L$ is the action of the unperturbed CFT:
\begin{equation} \label{AL}
\begin{split}
{\cal A}_{L}=\int \bigg\{ \frac{1}{4 \pi} \partial_z \varphi(z,\bar{z}) \, \partial_{\bar{z}} \varphi(z,\bar{z})
-\frac{ {\bf \mu}^2}{\sin \pi \beta^2} e^{-i \, \beta \, \varphi(z,\bar{z})} \bigg\} \, \frac{i \, dz \wedge d\bar{z}}{2}.
\end{split}
\end{equation}

The central charge of this CFT is
\begin{equation} \label{centLiou}
\begin{split}
c_L=1-6 \frac{\nu^2}{1-\nu}, \qquad \nu=1-\beta^2, 
\end{split}
\end{equation}
and its primary fields are parameterized by a continuous real parameter $\alpha$:
\begin{equation} \label{primLiou}
\begin{split}
\Phi_\alpha(z,\bar{z})=e^{\tfrac{i\, \alpha \beta \nu}{2 (1-\nu)}  \, \varphi(z,\bar{z})},
\end{split}
\end{equation}
with scaling dimensions $2 \Delta_\alpha$, where
\begin{equation} \label{Dalfa}
\begin{split}
\Delta_\alpha=\frac{\nu^2}{4(1-\nu)} \alpha \, (\alpha-2). 
\end{split}
\end{equation}
We use the same normalization for the fields as in \cite{Jimbo:2010jv}. 
Namely, the normalization is chosen such that the shift of $\alpha$ by the unit $2(1-\nu)/\nu$ 
corresponds, in the CFT language, to normal ordered multiplication by the primary field 
$e^{i\beta \varphi(z,\bar z)}$.

In this study, following \cite{Jimbo:2010jv}, we focus on the fundamental range $0 \leq \alpha < 2$. 
Extensions to a wider range of $\alpha$ can be achieved via an appropriate analytical continuation.


The primary fields (\ref{primLiou}) together with their descendants form a complete basis in the space of operators of the theory.  
Exploiting the fermionic basis, it was demonstrated in \cite{Jimbo:2010jv} that specific ratios of vacuum expectation values of these fields can be expressed in terms of a single function $\omega(\lambda,\mu|\alpha)$. More precisely, the expectation values are determined by a finite set of coefficients arising in the asymptotic expansions of $\omega(\lambda,\mu|\alpha)$ as $\lambda,\mu \to \pm \infty$: 
\begin{equation} \label{ommatr}
\begin{split}
\omega(\mu,\lambda|\alpha)\!=\!\!\!\sum\limits_{j,k=1}^{\infty} \!\!e^{-\epsilon_1 \mu \tfrac{2k-1}{\nu}} e^{-\epsilon_2 \lambda \tfrac{2j-1}{\nu}} 
\omega_{\epsilon_1 (2k-1),\epsilon_2(2j-1)}(\alpha),
\quad \! \epsilon_{1,2}\!=\!\pm, \quad \epsilon_1 \, \mu \!\to \!\infty, \quad \!
\epsilon_2 \, \lambda \! \to \! \infty.
\end{split}
\end{equation}
The coefficients define an infinite matrix $\omega_{2k-1,1-2j}(\alpha), \, j,k\in \mathbb{Z}.$ 
In this paper, we study its UV series expansion, for $j,k=1,\dots,\infty$, using an integrable description of the spectral problem in the UV CFT limit of the model, in order to determine the leading term in the UV expansion of vacuum expectation values. 

The significance of this matrix stems from the fact that ratios of vacuum expectation values of primary fields 
admit a direct representation in terms of its matrix elements \cite{Jimbo:2010jv}: 
\begin{equation}  \label{primVEV}
\begin{split}
\frac{\langle  \Phi_{\alpha+2 \,m\, \tfrac{1-\nu}{\nu}}(0) \rangle}{\langle  \Phi_{\alpha}(0) \rangle}=
\mu^{2 \, m \alpha-2 \, m^2\,(1-\tfrac{1}{\nu})}\, C_m(\alpha) \, \underset{1 \leq j,k \leq m }{\mbox{det}} \, \Omega_{kj}(\alpha), \qquad m=1,2,\dots
\end{split}
\end{equation}
where $C_m(\alpha)$ is given by
\begin{equation} \label{Cm}
\begin{split}
C_m(\alpha)=\prod\limits_{j=0}^{m-1} C_1(\alpha+2j\tfrac{1-\nu}{\nu}),
\end{split}
\end{equation}
with
\begin{equation} \label{C1}
\begin{split}
&C_1(\alpha)=i \, \nu \Gamma(\nu)^{4 x(\alpha)}\, \frac{\Gamma(-2 \nu x(\alpha))}{\Gamma(2 \nu x(\alpha))}\, \frac{\Gamma(x(\alpha))}{\Gamma(x(\alpha)+1/2)} \,
\frac{\Gamma(-x(\alpha)+1/2)}{\Gamma(-x(\alpha))} \cot(\pi x(\alpha)), \\
&\text{and} \qquad x(\alpha)=\frac{\alpha}{2}+\frac{1-\nu}{2 \nu}.
\end{split}
\end{equation}

The elements of the matrix $\Omega_{kj}(\alpha)$ are expressed in terms of $\omega_{2k-1,1-2j}(\alpha)$ as
\begin{equation}  \label{OM}
\begin{split}
\Omega_{kj}(\alpha)=\omega_{2k-1,1-2j}(\alpha)+\frac{i}{\nu} \, \delta_{j,k}\, 
\cot\left[ \frac{\pi}{2 \nu} (2k-1+\nu \, \alpha)\right], \qquad j,k=1,...,m.
\end{split}
\end{equation}


Finally, the parameter $\mu$ appearing in (\ref{primVEV}) is precisely the coupling constant in the sine-Gordon action (\ref{LSG}). It is related to the soliton mass ${\cal M}$ via \cite{Zamolodchikov:1995xk}:
\begin{equation} \label{muM}
\begin{split}
\mu={\cal M}^\nu \, \Pi(\nu)^\nu, \qquad \text{where} \qquad 
\Pi(\nu)=\frac{\sqrt{\pi}}{2} \frac{\Gamma\left(\tfrac{1}{2 \nu}\right)}{\Gamma\left(\tfrac{1-\nu}{2 \nu}\right)} \, \Gamma\left( \nu \right)^{-\tfrac{1}{\nu}}.
\end{split}
\end{equation}
Thus, all quantities appearing in (\ref{primVEV}) have now been specified, with the exception of  $\omega_{2k-1,1-2j}(\alpha)$, which we now turn to. 
The entire volume dependence of the expectation-value ratios in (\ref{primVEV}) is encoded in these matrix elements. In the infinite-volume limit, where $\omega_{2k-1,1-2j}(\alpha) \to 0,$ (\ref{primVEV}) is in agreement with the Lukyanov-Zamolodchikov formula \cite{Lukyanov:1996jj}. 

We now remove trivial constant factors from the original definition given  in \cite{Jimbo:2010jv}:
\begin{equation}  \label{omtom}
\begin{split}
\omega_{2k-1,1-2j}(\alpha)=\frac{i}{\pi \, \nu}\, \tilde{\omega}_{2k-1,1-2j}(\alpha), \qquad j,k \in \mathbb{Z}.
\end{split}
\end{equation}
The nontrivial content of the construction is encoded in $\tilde{\omega}_{2k-1,1-2j}(\alpha),$ which is defined through an integral representation involving functions that arise from the integrable description of the finite-volume ground-state 
energy:
\begin{equation} \label{omegatilde}
\begin{split}
\tilde{\omega}_{2k-1,1-2j}(\alpha)=\int\limits_{-\infty}^\infty dx \, e^{(2k-1)x} \, {\mathfrak m}(x) \, {\cal G}^{(\alpha)}_{1-2j}(x), 
\qquad j,k\in \mathbb{Z}
\end{split}
\end{equation}
Here, ${\mathfrak m}(x)$ is a measure constructed from the counting function $Z(x)$ of the nonlinear integral equation (NLIE) (\ref{DDV}) governing the finite-volume ground-state energy:
\begin{equation} \label{m}
\begin{split}
{\mathfrak m}(x)\equiv{\mathfrak m}[Z(x)]={\cal L}_+[Z(x+i\, 0)]+{\cal L}_-[Z(x-i\, 0)], \qquad \text{with} \qquad
{\cal L}_\pm[Z]=\frac{e^{\pm i \, Z}}{1+e^{\pm i \, Z}}.
\end{split}
\end{equation}
The functions ${\cal G}_{1-2j}^{(\alpha)}(x)$ satisfy the linear integral equation:
\begin{equation} \label{G1m2j}
\begin{split}
{\cal G}^{(\alpha)}_{1-2j}(x)-\int\limits_{-\infty}^\infty dy \, G_{\alpha}(x-y) \, {\mathfrak m}(y) \, {\cal G}^{(\alpha)}_{1-2j}(y)=e^{(1-2j)x}, \qquad j\in \mathbb{Z},
\end{split}
\end{equation}
where $G_{\alpha}(x)$ denotes the deformed kernel of the NLIE\footnote{Its relation to $R(\lambda|\alpha)$ in \cite{Jimbo:2010jv} is $G_{\alpha}(x)=-\nu \, R(\nu \, x|\alpha)$ with $\nu=1/(p+1)$.}. It is defined 
as the Fourier-transform
\begin{equation} \label{Galpha}
\begin{split}
G_{\alpha}(x)=\int\limits_{-\infty}^\infty \frac{d\omega}{2 \pi} \, \tilde{G}_{\alpha}(\omega) \, e^{i \, \omega \, x}, \qquad 
\text{with} \qquad 
\tilde{G}_{\alpha}(\omega)=\frac{ \sinh\left(\tfrac{\pi \omega (p-1)}{2}+i \tfrac{\pi \, \alpha}{2}\right) }
{2 \, \cosh\left(\tfrac{\pi \omega}{2}\right) \, \sinh\left(\tfrac{\pi \omega p}{2}+i \tfrac{\pi \, \alpha}{2}\right)}.
\end{split}
\end{equation}
Here $0<\al<2.$ Following standard NLIE conventions \cite{Feverati:1998dt,Feverati:1998uz,Feverati:1999sr,Feverati:2000xa}, we introduce $p$ as an alternative parameterization of $\nu$:
\begin{equation}  \label{pnu}
\begin{split}
\nu=\frac{1}{1+p}, \qquad p=1-\frac{1}{\nu}.
\end{split}
\end{equation}
In this parameterization, $0<p<1$ corresponds to the attractive regime, while $1<p<\infty$ describes 
the repulsive regime.

The parameter $\alpha$ in (\ref{Galpha}) acts as a deformation parameter: setting $\alpha=0$ recovers the kernel of the original NLIE (\ref{DDV}), denoted simply by $G(x)$.


The counting function $Z(x)$ is periodic with period $i \pi (p+1)$. In the fundamental strip $|\mathrm{Im}(x)|<\min(\pi,p \pi)$, it satisfies the nonlinear integral equation (NLIE) \cite{Destri:1992qk,Destri:1994bv}:
\begin{equation} \label{DDV}
\begin{split}
Z(x)=\ell \, \sinh x+\alpha_z+\frac{1}{i} \!\!\! \int\limits_{-\infty}^\infty dy  \left\{ G(x-y-i \, 0) \, L_+(y+i \, 0)-  G(x-y+i \, 0) \, L_-(y-i \, 0) \right\},
\end{split}
\end{equation}
where we introduced the shorthand notations:
\begin{equation} \label{Lpmdef}
\begin{split}
L_\pm(x)=\log\left(1+e^{\pm i \, Z(x)} \right), \qquad \ell={\cal M} \, L,
\end{split}
\end{equation}
with ${\cal M}$ and $L$ denoting the infinite-volume soliton mass and the finite spatial volume, respectively.

The counting function $Z(x)$ plays a central role in the integrable formulation of the spectral problem, as it allows one to compute the finite-volume ground-state energy of the sine-Gordon model:
\footnote{For the extension to the full spectrum, see \cite{Fioravanti:1996rz,Feverati:1998dt,Feverati:1998uz,Feverati:2000xa}.}
\begin{equation}
E_0(L)=-\frac{\cal M}{2 \, \pi \, i}\, \int\limits_{-\infty}^\infty \!\! dx \, \sinh x \, \log \frac{1+e^{i \, Z(x+i\, 0)}}{1+e^{-i \, Z(x-i\, 0)}}.
\end{equation}

We also include a twist parameter $\alpha_z$ in the NLIE, which allows us to discuss vacuum states of twisted sine-Gordon theories \cite{Fioravanti:1996rz,Feverati:1999sr,Feverati:2000xa}. This parameter becomes relevant when describing either minimal models and their integrable perturbations, or states in the underlying complex Liouville CFT. It is important, however, to distinguish between 
the two deformation parameters introduced: $\alpha$ characterizes the operator whose vacuum expectation value is being studied, while $\alpha_z$ induces a twist in the NLIE governing the finite-volume energy of the sandwiching state.

In \cite{enjhep}, it was argued that the quantity (\ref{omegatilde}) has the following UV expansion:
\begin{equation} \label{UVexpansion}
\begin{split}
\tilde{\omega}_{2k-1,1-2j}(\alpha)=\left(\frac{2}{\ell}\right)^{2(k+j-1)+\frac{2 (\alpha-2)}{p+1}} \, 
\sum\limits_{n=1}^\infty \tilde{\omega}_{2k-1,1-2j|n}^{(\alpha)}\, \left(\frac{\ell}{2}\right)^{\frac{4 (n-1)}{p+1}}.
\end{split}
\end{equation}
The expansion coefficients appearing in this expression are highly nontrivial and, in general consist of many contributions. In this work, we do not  aim at their full determination. 
Instead, our goal is to isolate the subset of terms that determine the leading UV behaviour of the ratios (\ref{primVEV})  of vacuum expectation values of primary fields, and to express them in terms of functions characterizing the ultraviolet limit of the model. More precisely, we focus on those contributions in the UV expansion that survive in the leading UV limit of the vacuum expectation values. To this end, we first analyse the UV behaviour of $Z(x)$ in more detail.


\section{The UV behaviour of the counting-function } \label{3}

The counting-function $Z(x)$ is the solution of the NLIE (\ref{DDV}). 
It governs the $\ell$-dependence of all quantities of interest, since 
through (\ref{m}) it determines the integration measure, which in turn introduces the $\ell$-dependence 
into the linear problem (\ref{G1m2j}), as well as into the defining expression for $\tom_{2k-1,1-2j}(\alpha)$ 
in (\ref{omegatilde}). 

Consequently, a consistent description of the UV behaviour of the expectation values requires 
the knowledge of the UV behaviour of the counting function. In this section, we address this problem. 

The leading-order qualitative UV behaviour of $Z(x)$ is well known from the literature 
\cite{Destri:1994bv,Destri:1997yz,Feverati:1998dt,Feverati:1998uz}. It can be described in terms of three different functions corresponding to three distinct regimes in $x$. Namely,
\begin{equation}  \label{ZUV}
\begin{split}
Z(x) \, x \simeq \left\{\begin{array}{ll} 
Z_+(x-\ltl) & \mbox{ for } \ltl \lesssim x, \\
z_0 & \mbox{ for } -\ltl \lesssim x \lesssim \ltl, \\
Z_-(x+\ltl) & \mbox{ for } x \lesssim-\ltl.
\end{array} \right.
\end{split}
\end{equation}

In the central regime, $Z(x)$ is well approximated by a constant $z_0,$ which is the $\ell \to 0$ limit 
of the counting function:
\begin{equation}  \label{z0def}
\begin{split}
z_0=\lim\limits_{\ell \to 0} Z(x).
\end{split}
\end{equation}

In the two outer regimes, the kink functions $Z_\pm(x)$ are defined as appropriate scaling limits of $Z(x)$:
\begin{equation}  \label{Zpmdef}
\begin{split}
Z_\pm(x)=\lim\limits_{\ell \to 0} Z(x\pm \ltl).
\end{split}
\end{equation}

These functions play a central role in the integrable description of the CFT arising in the UV limit of the sine-Gordon model, since the spectral data of the CFT $(c,\Delta_\pm)$ associated with the ground state can be expressed in terms of them as
\begin{equation}  \label{EpmCFT}
\begin{split}
E_0^{CFT}(L)&=E_0^+(L)+E_0^-(L), \\
E_0^\pm(L)&=\frac{2 \pi}{L} \, \left(\Delta_\pm-\frac{c}{24} \right)=\mp \frac{1}{ L} \intinf \!\! \frac{dx}{4 \pi i } \, e^{\pm x} 
\log \frac{1+e^{i \, Z_\pm(x+i \, 0)}}{1+e^{-i \, Z_\pm(x-i \, 0)}}.
\end{split}
\end{equation}

Standard UV analysis techniques \cite{Destri:1994bv,Destri:1997yz,Fioravanti:1996rz,Feverati:1998dt}, based on the kink equations (\ref{kinksol}), lead to the following expressions for the chiral energies:
\begin{equation}  \label{EpmCFTcomputed}
\begin{split}
E_0^\pm(L)&=-\frac{\pi}{12 \, L} \, \left(1-6\, (1-\nu)\, \left(\frac{\al_z}{\pi}\right)^2 \right).
\end{split}
\end{equation}

In the complex Liouville CFT formulation, these correspond to conformal dimensions of the form
\begin{equation}  \label{DLCFT}
\begin{split}
\Delta_\pm \to &\Delta_{1\pm \kappa_z}, \qquad \text{with} \quad \kappa_z=\frac{\al_z}{\pi} \, \frac{1-\nu}{\nu},
\end{split}
\end{equation}
where $\Delta_{1\pm \kappa_z}$ is computed from (\ref{Dalfa}).

Using (\ref{DDV}) and (\ref{Zpmdef}), one can derive the following nonlinear integral equations for $Z_\pm(x)$:
\begin{equation} \label{kinksol}
\begin{split}
Z_\pm(x)=\pm e^{\pm x}+\alpha_z+\frac{1}{i} \!\!\! \int\limits_{-\infty}^\infty dy  
\left\{ G(x-y-i \, 0) \, L^{(\pm)}_+(y+i \, 0)
-  G(x-y+i \, 0) \, L^{(\pm)}_-(y-i \, 0) \right\},
\end{split}
\end{equation}
where we introduced the shorthand notation
\begin{equation} \label{Lpmdef}
\begin{split}
L^{(\sigma)}_\pm(x)=\log\left(1+e^{\pm i \, Z_\sigma(x)} \right), \qquad \sigma \in \{\pm\}.
\end{split}
\end{equation}

These equations imply that the plateau value $z_0$ is related to the asymptotic behaviour of $Z_\pm(x)$ at $\mp\infty$:
\begin{equation}  \label{Zpm0}
\begin{split}
Z_{\pm}(\mp \infty)=z_0.
\end{split}
\end{equation}
The constant $z_0$ satisfies the so-called plateau equation \cite{Destri:1997yz}:
\begin{equation}  \label{plateq}
\begin{split}
z_0=\alpha_z+\frac{1}{i} \left( \intinf \! dx \, G(x) \right) \,
\left( \log\left( 1+e^{i  \, z_0} \right)-\log\left( 1+e^{-i \, z_0} \right) \right).
\end{split}
\end{equation}
In the absence of special objects and multi-plateau structures \cite{Destri:1997yz,Feverati:1998dt,Feverati:1998uz,Feverati:1999sr,Feverati:2000xa},  
it takes the simple form
\begin{equation}  \label{z0value}
\begin{split}
z_0=\frac{2\, p\, \alpha_z}{p+1}.
\end{split}
\end{equation}

In \cite{Bajnok:2022ucr} an asymptotic solution of (\ref{DDV}) was introduced in order to unify the behaviour in the three regimes within a single function, while preserving the correct leading UV behaviour. 
Its definition is
\begin{equation}  \label{Zasdef}
\begin{split}
Z_{as}(x)=Z_+(x-\ltl)+Z_-(x+\ltl)-z_0.
\end{split}
\end{equation}
In \cite{Bajnok:2022ucr} it was proposed\footnote{In \cite{Bajnok:2022ucr} the construction was formulated for Thermodynamic Bethe Ansatz (TBA) systems; the same reasoning applies to the NLIE case without modification.}
 that the small-$\ell$ expansion of physical observables can be obtained 
 by a systematic expansion around this solution.

Later, combining insights from perturbed conformal field theory (PCFT) arguments with the analysis of 
high-precision numerical solutions of the NLIE (\ref{DDV}), the following Ansatz was conjectured 
in \cite{enjhep} for the small-$\ell$ expansion of $Z(x)$:
\begin{equation}  \label{ZUVansatz}
\begin{split}
Z(x)&=Z_{as}(x)+\de Z(x), \\
\de Z(x)&= \sum\limits_{n=1}^{\infty} \left(\frac{\ell}{2}\right)^{\tfrac{4 \, n}{p+1}} \, 
\left( \de Z^{(n)}_{+}(x-\ltl)+\de Z^{(n)}_{-}(x+\ltl)-\de Z^{(n)}_0 \right),
\end{split}
\end{equation}
where $\de Z^{(n)}_{\pm}(x)$ denotes the $n$th UV correction to the kink functions, while $\de Z^{(n)}_0$ is the corresponding plateau value:
\begin{equation}  \label{dZn0}
\begin{split}
\de Z^{(n)}_0=\de Z^{(n)}_{\pm}(\mp \infty).
\end{split}
\end{equation}
All quantities $\de Z^{(n)}_\pm(x)$ and $\de Z^{(n)}_0$ are independent of $\ell$.
Thus, (\ref{ZUVansatz}) suggests that deviations from the asymptotic solution can be organized as a series 
running in integer powers of $\ell^{\tfrac{4}{p+1}}$, with each order preserving the same kink-plateau structure 
as the asymptotic solution. 
The higher-order corrections $\de Z^{(n)}_\pm(x)$ and $\de Z^{(n)}_0$ contribute only to higher-order terms in the UV expansion (\ref{UVexpansion}) of $\tom_{2k-1,1-2j}(\alpha)$. For instance, $\de Z^{(1)}_\pm(x)$ and $\de Z^{(1)}_0$ enter  only at the second-order level in the UV expansion of $\tom_{2k-1,1-2j}(\alpha)$.



\section{UV predictions from field theory } \label{4}

Viewing the sine-Gordon model as a perturbation of complex Liouville theory, one can use the Liouville three-point functions \cite{Dotsenko:1984nm,Zamolodchikov:1995aa,Teschner:1995yf} to obtain predictions for the leading UV behaviour of ratios of expectation values of primary fields (\ref{primLiou}). Applied to the quantity of interest (\ref{primVEV}), this leads to the following expected UV asymptotics \cite{Jimbo:2010jv}:
\begin{equation} \label{VEVcft}
\begin{split}
\frac{\langle  \Phi_{\alpha+2 \,m\, \tfrac{1-\nu}{\nu}}(0) \rangle}{\langle  \Phi_{\alpha}(0) \rangle}\simeq 
\left(\frac{2 \, \pi}{ L} \right)^{2(\Delta_{\al+2 m (1-\nu)/\nu}-\Delta_{\al})} \, 
\prod\limits_{j=0}^{m-1}\, {\cal V}(\alpha+ 2\,j \, \tfrac{1-\nu}{\nu},\kappa_\Delta),
\end{split}
\end{equation}
where $\Delta$ denotes the conformal dimension of the sandwiching Liouville primary field, and $\kappa_\Delta$ is defined as
\begin{equation}  \label{kappaDelta}
\begin{split}
\kappa_\Delta=\sqrt{1-\frac{\Delta}{\Delta_1}}, \qquad \text{with} \quad \Delta_1=-\frac{\nu^2}{4(1-\nu)}.
\end{split}
\end{equation}

We note that, in order to obtain  expectation values with $\Delta$ governed by the solution of (\ref{DDV}), 
(\ref{VEVcft}) must be evaluated at the points $\Delta \to \Delta_{1\pm \kappa_z}$, where $\kappa_z$ is given in (\ref{DLCFT}).

The functions entering (\ref{VEVcft}) are defined as follows \cite{Jimbo:2010jv}:
\begin{equation} \label{prim2CFT}
\begin{split}
{\cal V}(\alpha,\kappa)=\mu^2 \, \Gamma(\nu)^2 \, Y(x(\alpha)) \, W(\alpha,\kappa) \, W(\alpha,-\kappa),
\end{split}
\end{equation}
where $\mu$ and $x(\alpha)$ are defined in (\ref{muM}) and (\ref{C1}), respectively, and
\begin{equation} \label{Yx}
\begin{split}
Y(x)&=-2 \, \nu\, x \cdot \frac{\Gamma^2(\nu x+1/2-\nu/2) \, \Gamma(\nu-2 \nu x)}{\Gamma^2(1/2+\nu/2-\nu x) \, \Gamma(2 \nu x+1-\nu)}\cdot 
\frac{\Gamma(-2 \nu x)}{\Gamma(2 \nu x)}, \\
W(\alpha,\kappa)&=\frac{\Gamma(\alpha\nu/2-\nu+1+\kappa \nu)}{\Gamma(-\alpha \nu/2+\nu+\kappa\nu)}.
\end{split}
\end{equation}

Using the explicit expressions for the conformal dimensions and (\ref{VEVcft}), the leading volume dependence of the expectation values (\ref{primVEV}) takes the form in the UV limit
\begin{eqnarray} \label{simVEV}
\frac{\langle  \Phi_{\alpha+2 \,m\, \tfrac{1-\nu}{\nu}}(0) \rangle}{\langle  \Phi_{\alpha}(0) \rangle}\!\!
&=&
{\cal C}(\alpha,m,\nu) \, L^{-2\, m \left(m-(m+1)\nu \right)-2\, m \al \, \nu}\,
(1+O(\ell^{4 \, \nu})), \quad \!m\!=\!1,2,... \\
\!\! \mbox{with:} \quad {\cal C}(\alpha,m,\nu) \!\!&=&\!\! 
\left({2 \, \pi} \right)^{2(\Delta_{\al+2 m (1-\nu)/\nu}-\Delta_{\al})} \, 
\prod\limits_{j=0}^{m-1}\, {\cal V}(\alpha+ 2\,j \, \tfrac{1-\nu}{\nu},\kappa_\Delta). \label{simVEV1}
\end{eqnarray}
The coefficient ${\cal C}(\alpha,m,\nu)$ governs the leading ultraviolet contribution 
and can be expressed in terms of known three-point functions in the complex Liouville CFT defined on the cylinder. 
More precisely, we obtain \cite{Jimbo:2010jv}:
\begin{equation} \label{3ptL}
\begin{split}
{\cal C}(\alpha,m,\nu)=\left({2 \, \pi} \right)^{2(\Delta_{\al+2 m (1-\nu)/\nu}-\Delta_{\al})} \, 
\frac{\langle \Phi_{1-\kappa_{\Delta}}(-\infty) \Phi_{\alpha+2 \,m\, \tfrac{1-\nu}{\nu}}(0) \Phi_{1+\kappa_{\Delta}}(\infty)\rangle_{CFT}}{\langle \Phi_{1-\kappa_{\Delta}}(-\infty) \Phi_{\alpha}(0) \Phi_{1+\kappa_{\Delta}}(\infty) \rangle_{CFT}},
\end{split}
\end{equation}
where the cylinder coordinates $z=-\infty, 0, +\infty$ correspond, under the conformal map $w=e^{-z}$, to the standard insertion points $w=\infty, 1, 0$ on the Riemann sphere. 
The determination of this coefficient within the integrable framework is the 
central objective of this work. 

This requires the analysis of the the UV expansion (\ref{UVexpansion}) of $\tom_{2k-1,1-2j}(\alpha)$. In \cite{enjhep} it was argued,  by inserting the expansion (\ref{UVexpansion}) into (\ref{primVEV}), 
that the computation of ${\cal C}(\alpha,m,\nu)$ in general involves the first $m(m-1)/2+1$ coefficients 
from the UV series (\ref{UVexpansion}). 
Due to the complexity of higher order terms, only the first two coefficients were explicitly computed in \cite{enjhep}, which allowed the cases $m=1$ and $m=2$ to be treated. In particular, explicit expressions were obtained for ${\cal C}(\alpha,1,\nu)$ and ${\cal C}(0,2,\nu)$, the latter requiring a rather involved computation.

In the present work, we significantly refine this picture. First, we show that the computation of ${\cal C}(\alpha,m,\nu)$ does not require all $m(m-1)/2+1$ coefficients of (\ref{UVexpansion}); in fact, the first $m$ coefficients are sufficient. Second, we argue that even within these, only a restricted and relatively simple 
subset of contributions actually enters the leading coefficient ${\cal C}(\alpha,m,\nu).$

\section{The leading UV term in the vacuum expectation values from integrability } \label{5}

In order to formulate the leading UV expression for vacuum expectation values, several key quantities derived from the kink functions must first be introduced.


Following from the definitions (\ref{Zpmdef}), the kink functions inherit the same periodicity property as the counting function. In particular, they are periodic with period $i \,\pi\, (p+1)$. As a consequence, they admit a large-argument expansion in the plateau regime of the form
\begin{equation} \label{kinkplat}
\begin{split}
Z_-(x)&\stackrel{x\to +\infty}{=}z_0+ \sum\limits_{n=1}^{\infty} c_n \, e^{-\tfrac{2 \,n\,  x}{p+1}},  \\
Z_+(x)&\stackrel{x \to -\infty}{=}z_0+ \sum\limits_{n=1}^{\infty} \tilde{c}_n \, e^{\tfrac{2 \, n\,  x}{p+1}}.
\end{split}
\end{equation}
The coefficients $c_n$ and $\tilde{c}_n$ are related to the $n$th non-local conserved charges of the theory \cite{Bazhanov:1996dr,Bazhanov:1996aq}.

The terms appearing in the systematic expansion (\ref{UVexpansion}) are constructed from solutions of linear integral equations defined on kink backgrounds \cite{enjhep}. To this end, we introduce the functions ${\cal G}_{-(2j-1)}^{(\alpha)(-)}(x)$ and ${\cal G}_{+(2k-1)}^{(-\alpha)(+)}(x)$ as the solutions of the following linear equations:
\begin{equation} \label{kinkmalpha}
\begin{split}
{\cal G}_{-(2j-1)}^{(\alpha)(-)}(x)
-\int\limits_{-\infty}^\infty \!\! dy \,
G_{\alpha}(x-y)\,
{\mathfrak m}_-(y)\,
{\cal G}_{-(2j-1)}^{(\alpha)(-)}(y)
=
e^{(1-2j)x},
\qquad j=1,2,\ldots
\end{split}
\end{equation}
and
\begin{equation} \label{kinkpalpha}
\begin{split}
{\cal G}_{+(2k-1)}^{(-\alpha)(+)}(x)
-\int\limits_{-\infty}^\infty \!\! dy \,
G_{-\alpha}(x-y)\,
{\mathfrak m}_+(y)\,
{\cal G}_{+(2k-1)}^{(-\alpha)(+)}(y)
=
e^{(2k-1)x},
\qquad k=1,2,\ldots
\end{split}
\end{equation}
where, for notational convenience, we introduced the shorthand notation
${\mathfrak m}_\pm(x)={\mathfrak m}[Z_\pm(x)]$
for the kink measures{\footnote{Here, $G_{-\al}(x)=G_\al(-x),$ which is equivalent to taking (\ref{Galpha}) at 
$\al \to -\al.$ }}.

It is important to understand the large-$x$ asymptotic behaviour of these functions. In the two regimes, they take the form
\begin{equation} \label{scGmasy}
\begin{split}
{\cal G}_{-(2j-1)}^{(\alpha)(-)}(x)\,\, & \stackrel{x \to +\infty}{=}e^{\tfrac{\alpha \,x}{p+1}}\, \sum\limits_{n=1}^\infty c_n^{(-(2j-1),\alpha)} \, e^{-\tfrac{2 n x}{p+1}}, \\
{\cal G}_{-(2j-1)}^{(\alpha)(-)}(x)-e^{-(2j-1)\, x} \,\, &\stackrel{x \to -\infty}{=} \sum\limits_{n=1-\Theta(\alpha)}^\infty a_n^{(-(2j-1),\alpha)} \, e^{\tfrac{\alpha+2 n}{p} x}+
 \sum\limits_{n=0}^\infty b_n^{(-(2j-1),\alpha)} \, e^{(1+2 n)\,x},
\end{split}
\end{equation}
where $\Theta(\alpha)$ is the Heaviside step function, so that $1-\Theta(\alpha)=1$ for $\alpha=0$ and $0$ for $\alpha>0$.

Completely analogous expressions hold in the "$+$"-kink  sector:
\begin{equation} \label{scGpasy}
\begin{split}
{\cal G}_{+(2k-1)}^{(-\alpha)(+)}(x)\,\, & \stackrel{x \to -\infty}{=}e^{-\tfrac{\alpha \,x}{p+1}}\, \sum\limits_{n=1}^\infty c_n^{(+(2k-1),-\alpha)} \, e^{\tfrac{2 n x}{p+1}}, \\
{\cal G}_{+(2k-1)}^{(-\alpha)(+)}(x)-e^{(2k-1)\, x} \,\, &\stackrel{x \to \infty}{=} \! \! \! \! \! \sum\limits_{n=1-\Theta(\alpha)}^\infty \!\!\!\! a_n^{(+(2k-1),-\alpha)} \, e^{-\tfrac{\alpha+2 n}{p} x}+
 \sum\limits_{n=0}^\infty b_n^{(+(2k-1),-\alpha)} \, e^{-(1+2 n)\,x}.
\end{split}
\end{equation}

The first lines in (\ref{scGmasy}) and (\ref{scGpasy}) can be supported by high-precision numerical solutions of (\ref{kinkmalpha}) and (\ref{kinkpalpha}), while the second lines follow directly from the large-argument expansion of the kernel $G_{\al}(x)$ given in (\ref{Gsoral}).

At this point, all ingredients required for the leading-order UV formula of vacuum expectation values have been introduced.


In \cite{enjhep}, the leading UV term was computed for the expectation values
$\tfrac{\langle \Phi_{\alpha+2 \, \tfrac{1-\nu}{\nu}}(0)\rangle}{\langle \Phi_{\alpha}(0)\rangle}$ and $\langle \Phi_{4 \, \tfrac{1-\nu}{\nu}}(0)\rangle$. This was achieved by determining the first two coefficients in the series (\ref{UVexpansion}) for $\tom_{2k-1,1-2j}(\alpha)$ using integrability techniques. Although the final expression for the ratio $\tfrac{\langle \Phi_{\alpha+2 \, \tfrac{1-\nu}{\nu}}(0)\rangle}{\langle \Phi_{\alpha}(0)\rangle}$ turned out to be relatively simple, the result for $\langle \Phi_{4 \, \tfrac{1-\nu}{\nu}}(0)\rangle$ was considerably more involved and did not reveal a clear underlying structure.

In this paper, mainly based on high-precision numerical work, we show that there is in fact a clear
structure, and a final integrability result for the leading coefficient ${\cal C}(\alpha,m,\nu)$
entering (\ref{simVEV}) for arbitrary values of $m$, and not only for the cases $m=1,2$ discussed in \cite{enjhep}.
The key observation is that, although the $n$th term in the series (\ref{UVexpansion}) is composed of a large number of contributions, only a very specific subset of these terms is relevant from the point of view of the leading UV formula for expectation values. To make this precise, we introduce the
effective $\tom_{2k-1,1-2j}^{eff}(\alpha)$ matrix. Here, "effective" means that it contains only those contributions at each order of the series (\ref{UVexpansion}) which actually contribute to the leading-order coefficient ${\cal C}(\alpha,m,\nu)$. With this splitting, we write at each order
\begin{equation}
\begin{split} \label{tomn}
\tom_{2k-1,1-2j|n}(\alpha)=\tom_{2k-1,1-2j|n}^{eff}(\alpha)+\tom_{2k-1,1-2j|n}^{rest}(\alpha), \quad n=1,2,...,
\end{split}
\end{equation}
where $\tom_{2k-1,1-2j|n}^{rest}(\alpha)$ denotes all contributions that do not contribute to ${\cal C}(\alpha,m,\nu)$.
Based on high-precision numerical computations, we conjecture the following simple form for the effective part:

\begin{equation} \label{tomeffn}
\begin{split}
\tom_{2k-1,1-2j|n}^{eff}(\alpha)=c_n^{(-(2j-1),\al)}\,I_n^{(k)} , \quad n=1,2,...,
\end{split}
\end{equation}
where $c_n^{(-(2j-1),\al)}$ is the $n$th coefficient in the plateau expansion of 
${\cal G}_{1-2j}^{(\al),(-)}(x)$ in (\ref{scGmasy}), 
and $I_n^{(k)}$ is defined by
\begin{equation} \label{InkCP}
\begin{split}
I_{n}^{(k)}&=I_{n,0}^{(k)}+I_{n,1}^{(k)}, \\
I_{n,0}^{(k)}&=\CP\limits_{\ell \to 0} \int\limits_{-\log\tfrac{2}{\ell}}^\infty dx \,\, e^{\left(2k-1+\tfrac{\al}{p+1}
-\tfrac{2 n}{p+1} \right)\, x} \, {\mathfrak m}_+(x),  \\
I_{n,1}^{(k)}&=\CP\limits_{\ell \to 0} \int\limits_{-\log\tfrac{2}{\ell}}^\infty dx \,\, 
e^{\left(\tfrac{\al}{p+1}-\tfrac{2 n}{p+1} \right)\, x} \,
\left[ {\cal G}_{2k-1}^{(-\al)(+)}(x)-e^{(2k-1)x} \right]
({\mathfrak m}_+(x)-{\mathfrak m}_0).
\end{split}
\end{equation}
Here we introduced the symbolic operator $\CP\limits_{\ell \to 0}$, which denotes the extraction of the constant term in the $\ell \to 0$ expansion.

For the case $0<\alpha$, explicit integral representations can be given for $I_{n,0}^{(k)}$ and $I_{n,1}^{(k)}$. They take the form
\begin{equation} \label{In0k}
\begin{split}
I_{n,0}^{(k)}&= \int\limits_{-\infty}^\infty dx \,\, e^{\left(2k-1+\tfrac{\al}{p+1}
-\tfrac{2 n}{p+1} \right)\, x} \, {\mathfrak m}_+^{reg(n,k|\al)}(x), \qquad \mbox{where:} \\
{\mathfrak m}_+^{reg(n,k|\al)}(x)&= {\mathfrak m}_+(x)-\!\!\!
\sum\limits_{s=0}^{n-n_{min}^{(k,\al)}-1} {\mathfrak m}^+_s \,\, e^{\tfrac{2 s}{p+1} \,x}, \qquad \mbox{with:} \\
n^{(k,\al)}_{min}&=\left[ \frac{(p+1)(2k-1)}{2}+\frac{\al}{2} \right],
\end{split}
\end{equation}
where $[..]$ denotes the integer part.

Moreover,
\begin{equation} \label{In1k}
\begin{split}
I_{n,1}^{(k)}=\int\limits_{-\infty}^\infty dx \,\, 
e^{\left(\tfrac{\al}{p+1}-\tfrac{2 n}{p+1} \right)\, x} \,
\left( {\cal G}_{2k-1}^{(-\al)(+)}(x)-e^{(2k-1)x} \right)
\left({\mathfrak m}_+(x)-{\mathfrak m}_0-\sum\limits_{s=1}^{n-1} {\mathfrak m}_s^+ \, e^{\tfrac{2\, s}{p+1} \, x} \right).
\end{split}
\end{equation}

Finally, the coefficients ${\mathfrak m}_s^+$ are those appearing in the plateau expansion of ${\mathfrak m}_+(x)$:
\begin{equation} \label{masympts}
\begin{split}
{\mathfrak m}_\pm(x)={\mathfrak m}[Z_\pm(x)] &\stackrel{x \to \mp \infty}{=}
{\mathfrak m}_0+\sum\limits_{n=1}^\infty \mm_n^{\pm} \, e^{\pm \tfrac{2 \, n \, x}{p+1}}, \qquad \mm_0=\mm[z_0].
\end{split}
\end{equation}
Using (\ref{m}), these coefficients can be expressed in terms of $c_s$ and $\tilde{c}_s$ from (\ref{kinkplat}). The subtleties related to the precise evaluation of these expressions, in particular the role of the $\pm\, i\,0$ prescriptions, are discussed at the end of the section.

Inserting the UV series representation (\ref{UVexpansion}) of $\tom_{2k-1,1-2j}^{eff}(\alpha)$ with coefficients given in (\ref{tomeffn}) into the expectation value formula (\ref{primVEV}), one obtains the final expression for the leading UV coefficient 
\begin{equation} \label{calCres}
\begin{split}
{\cal C}(\al,m,\nu)&=  \left(\tfrac{i {\cal M}^{2 \nu } }{\pi \,\nu } \right)^m 
\Pi(\nu)^{2 \, m \, \nu \, \alpha-2 \, m^2\,(\nu-1)} 
 C_m(\alpha) \\
&\times
 \, 2^{2 (\Delta_{\al+2 \, m\, (1-\nu)/\nu}-\Delta_\al)}
\underset{1 \leq s,j \leq m }{\mbox{det}} \! c_s^{(-(2j-1),\al)}
 \underset{1 \leq n,k \leq m }{\mbox{det}} \! I_n^{(k)}.
\end{split}
\end{equation}
Together with (\ref{3ptL}), this implies the following representation for 
the ratios of cylinder 3-point functions of the 
complex Liouville CFT:
\begin{equation} \label{calC3res}
\begin{split}
\frac{\langle \Phi_{1-\kappa_{\Delta}}(-\infty) \Phi_{\alpha+2 \,m\, \tfrac{1-\nu}{\nu}}(0) \Phi_{1+\kappa_{\Delta}}(\infty)\rangle_{CFT}}{\langle \Phi_{1-\kappa_{\Delta}}(-\infty) \Phi_{\alpha}(0) \Phi_{1+\kappa_{\Delta}}(\infty) \rangle_{CFT}}&=
\left(\tfrac{i \, \mu^{2}}{\pi \,\nu } \right)^m 
\Pi(\nu)^{2 \, m \, \nu \, (\alpha-1)-2 \, m^2\,(\nu-1)} 
 C_m(\alpha) \\
&\times 
\underset{1 \leq s,j \leq m }{\mbox{det}}  c_s^{(-(2j-1),\al)}
 \underset{1 \leq n,k \leq m }{\mbox{det}} \! I_n^{(k)}\!,
\end{split}
\end{equation}
where with the help of the mass gap relation (\ref{muM}), we expressed this ratio 
in terms of the dimensionful parameter $\mu$ of the complex Liouville theory (\ref{AL}). 
Here, $\Delta_\al$, $C_m(\alpha)$  and $\Pi(\nu)$ are defined in (\ref{Dalfa}), (\ref{Cm}) and (\ref{muM}), respectively, while the matrix elements $c_s^{(-(2j-1),\al)}$ and $I_n^{(k)}$ are given in (\ref{scGmasy}) and (\ref{InkCP}).

In deriving (\ref{calCres}), we used the identity satisfied by $\tom_{2k-1,1-2j}^{eff}(\alpha)$:
\begin{equation} \label{detid}
\begin{split}
&\lim_{\epsilon \to 0} \bigg\{ \epsilon^{-m(m-1)/2}  \underset{1 \leq j,k \leq m }{\mbox{det}} \!\!\left( \sum\limits_{n=1}^\infty   
\epsilon^{(n-1)}\, 2^{2(j+k-1)+(2 \al-4 \, n) \nu}  \, \tom_{2k-1,1-2j|n}^{eff}(\alpha)  \right) \bigg\}
\!\! \\
&=\, 2^{2 (\Delta_{\al+2 \, m\, (1-\nu)/\nu}-\Delta_\al)}  \underset{1 \leq s,j \leq m }{\mbox{det}} \!( c_s^{(-(2j-1),\al)}) 
 \underset{1 \leq n,k \leq m }{\mbox{det}} \!( I_n^{(k)}).
\end{split}
\end{equation}

Formula (\ref{calCres}) is the main result of this work. We have verified it numerically 
in a variety of cases, exploring both the  attractive and repulsive regimes of the theory, 
and scanning over a wide range of values of all three parameters $\al$, the coupling constant $p$, 
and the twist $\alpha_z$, up to $m=4$. 
In all cases considered, we found excellent agreement with the analytic prediction (\ref{simVEV1}).

Appendix \ref{appC} contains numerical data for a selection of representative cases.

In Appendix \ref{appA}, we show how the effective contribution (\ref{tomeffn}) arises as part of the $n$th-order term in the small-volume expansion of $\tom_{2k-1,1-2j}(\alpha)$. This demonstrates that the effective part is not an artificially constructed contribution, but a genuinely present component of the UV expansion. In practice, we isolate the contributions that form the effective sector and show numerically that only these terms contribute to the leading UV behaviour of the expectation values.

We close this section by clarifying the proper treatment of the $\pm\, i\,0$ prescriptions entering our formulas through the measure $\mm(x)$ defined in (\ref{m}). It is straightforward to see that, at infinity in the plateau regimes, the integration contour can be deformed to the real axis, which, in view of (\ref{m}), implies
\begin{equation} \label{mek}
\begin{split}
\mm_0=1, \qquad \mm_j^{\pm}=0, \quad j=1,2,\dots
\end{split}
\end{equation}

However, from (\ref{m}) it is also clear that the correct implementation of the $\pm i\,0$ prescriptions in $\mm_\pm(x)$ requires the replacement
\[
\mm_\pm(x) \;\to\; {\cal L}_+[Z_\pm(x+i\,0)] + {\cal L}_-[Z_\pm(x-i\,0)]
\]
in any formula. The corresponding plateau expansions of the ${\cal L}_\pm[Z_\pm(x)]$ functions read
\begin{equation} \label{calLsor}
\begin{split}
{\cal L}_\sigma[Z_\pm(x)] &\stackrel{x \to \mp \infty}{=} 
{\cal L}_{\sigma}[z_0]+\sum\limits_{n=1}^\infty {\cal L}^{(\pm)}_{\sigma,n} \, e^{\pm \tfrac{2 \, n \, x}{p+1}}, \qquad \sigma \in \{\pm \}. 
\end{split}
\end{equation}
Using formulas (\ref{m}) and (\ref{kinkplat}), the coefficients can be expressed in a straightforward manner in terms of the coefficients $c_s$ and $\tilde{c}_s.$ 

With these conventions, a prototype integral for (\ref{In1k}) should be understood as follows:
\begin{equation} \label{protoIint}
\begin{split}
&\int\limits_{-\infty}^\infty dx \,\, 
f(x) \,\left({\mathfrak m}_+(x)-{\mathfrak m}_0-\sum\limits_{s=1}^{n-1} {\mathfrak m}_s^+ \, e^{\tfrac{2\, s}{p+1} \, x} \right) \\
&=\int\limits_{-\infty}^\infty dx \,\, 
f(x) \,\left({\cal L}_+[Z_+(x+i\, 0)]-{\cal L}_{+}[z_0]-\sum\limits_{s=1}^{n-1} {\cal L}^{(+)}_{+,s} \, e^{\tfrac{2  s  (x+i\, 0)}{p+1}} \right) \\
&\quad+\int\limits_{-\infty}^\infty dx \,\, 
f(x) \,\left({\cal L}_-[Z_+(x-i\, 0)]-{\cal L}_{-}[z_0]-\sum\limits_{s=1}^{n-1} {\cal L}^{(+)}_{-,s} \, e^{\tfrac{2  s  (x-i\, 0)}{p+1}} \right) \\
&=\int\limits_{-\infty}^\infty dx \,\, 
f(x+i \, \eta) \,\left({\cal L}_+[Z_+(x+i\, \eta)]-{\cal L}_{+}[z_0]-\sum\limits_{s=1}^{n-1} {\cal L}^{(+)}_{+,s} \, e^{\tfrac{2  s  (x+i\, \eta)}{p+1}} \right) \\
&\quad+\int\limits_{-\infty}^\infty dx \,\, 
f(x-i \, \eta) \,\left({\cal L}_-[Z_+(x-i\, \eta)]-{\cal L}_{-}[z_0]-\sum\limits_{s=1}^{n-1} {\cal L}^{(+)}_{-,s} \, e^{\tfrac{2  s  (x-i\, \eta)}{p+1}} \right),
\end{split}
\end{equation}
where $0<\eta$ is a small contour deformation parameter and $f(x)$ is assumed to be analytic in a neighbourhood of the real axis. In the concrete integrals arising in the UV analysis, the maximal admissible value of $\eta$ is constrained by the singularity structure of the kernel $G_\alpha(x)$. Typically, $\eta$ must be chosen smaller than half the distance between the real axis and the nearest singularity of $G_\alpha(x)$.

For integrals other than (\ref{protoIint}), the proper treatment of the $\pm i\, 0$ prescription is completely analogous. 

\section{Summary and discussion} \label{9}

In this paper, starting from an integrable formulation of expectation values of local fields in the sine-Gordon model \cite{Jimbo:2010jv}, 
we proposed an explicit formula (\ref{calCres}) for the leading term in the UV-limit expansion of ratios of vacuum expectation values of primary operators
\[
\frac{\langle \Phi_{\alpha+2 \, m \frac{1-\nu}{\nu}}(0) \rangle}{\langle \Phi_{\alpha}(0) \rangle}, \qquad m=1,2,..., \qquad 0\leq \al <2, 
\qquad 0<\nu<1, 
\] 
expressed in terms of kink functions that describe the UV limit of the model. 
The formula was obtained through analytical small-volume calculations and high-precision numerical analysis, 
which together led us to its explicit form.

The leading coefficient can be interpreted as a diagonal{\footnote{In this context the  term diagonal means 3-point functions with two coinciding operators. }} 
3-point function in the complex Liouville conformal field theory describing the UV limit of the sine-Gordon model. 
Importantly, unlike expressions written in terms of elementary or special functions, it does not admit an immediately evaluable closed-form expression. 
Instead, it specifies how the leading coefficient can be systematically obtained from the solutions of the kink NLIE, 
thereby establishing a concrete link between the integrable structure of the model and its conformal UV data.  

The results can be applied to concrete twisted NLIE cases with \(p \in \mathbb{N}\) \cite{Fioravanti:1996rz,Feverati:1999sr}, 
providing an integrable description of several diagonal 3-point couplings of primary fields in the unitary Virasoro minimal models \(Vir(p,p+1)\).  
At certain special twist values associated with these minimal models, the analytic structure of the solutions becomes significantly more intricate, 
and a complete analysis of these models remains a compelling challenge for future work.

There are two further interesting models in the literature for which, similarly to our case, an integrable formulation of the spectrum and expectation values is available: the $N=1$ supersymmetric (SUSY) sine-Gordon model \cite{Babenko:2019tvv} and the sinh-Gordon model \cite{Negro:2013wga,Negro:2013rxa,Bajnok:2019yik}.

Due to the periodicity properties of the functions entering the integrable
description of the spectral problem, we expect that our results can be extended to the $N=1$
SUSY sine-Gordon model in a nearly straightforward manner.

In contrast, as a consequence of the absence of such periodicity, our results cannot be applied to the
sinh-Gordon model in a straightforward way. Nevertheless, a relation analogous to our final formula
(\ref{calCres}) may still exist, providing an integrability-based description of the UV limit of expectation values in this model as well.

Overall, this work demonstrates that integrability provides not only a framework for highly precise numerical evaluation of physical quantities,
but also a systematic approach to extracting universal UV data directly from the kink NLIE. This highlights a deeper interplay between integrable quantum
field theories and their conformal ultraviolet limits, and suggests new avenues for understanding how UV information is encoded within integrable structures.


\vspace{1cm}
{\tt Acknowledgments}

\noindent 
The authors would like to thank to Zoltan Bajnok for useful discussions. 
The research reported in this paper was supported  by the National Research, Development and Innovation 
Office (NKFIH), Hungary, under the OTKA Grant NKKP-152467.

\appendix

\section{Extracting the terms defining $\tilde{\omega}_{2k-1,1-2j}^{eff}(\alpha)$ } \label{appA}

In this appendix, we consider the UV expansion (\ref{UVexpansion}) of $\tom_{2k-1,1-2j}(\al)$ and 
show how the terms $\tom_{2k-1,1-2j|n}^{eff}(\al)$ given in (\ref{tomeffn}) appear in the 
$n$th term of the full UV expansion of $\tom_{2k-1,1-2j}(\al)$. For the UV expansion, we use the 
method worked out in \cite{enjhep}, and we restrict our attention to contributions stemming from 
the asymptotic solution (\ref{Zasdef}), neglecting terms descending from corrections to this  
leading-order background. We do so because, based on physical intuition, one may expect that, at 
leading order, the asymptotic solution should fully describe the vacuum expectation values as well. 

Following \cite{enjhep} we give two simply related, equivalent formulations for $\tom_{2k-1,1-2j}(\al)$: 
\begin{equation} \label{om1m1def}
\begin{split}
\tom_{2k-1,1-2j}(\al)=\intinf \! dx \, e^{(2k-1)\,x}\, \mm(x) \, \scG_{1-2j}^{(\al)}(x)
=\intinf \! dx \, e^{(2k-1)\,x} \, \scG_{1-2j,m}^{(\al)}(x), \qquad j=1,2,,...
\end{split}
\end{equation}
where $\mm(x)\equiv\mm[Z(x)]$ is the integration measure (\ref{m}) related to the solution of 
the NLIE (\ref{DDV}), and 
$\scG_{1-2j}^{(\al)}(x)$ and $\scG_{1-2j,m}^{(\al)}(x)$ are two functions simply related to each other by:
\begin{equation} \label{scGvsGm}
\begin{split}
\scG_{1-2j,m}^{(\al)}(x)=\mm(x)\, \scG_{1-2j}^{(\al)}(x).
\end{split}
\end{equation}
They satisfy the following linear equations\footnote{The first equation is a repetition of 
(\ref{G1m2j}). }:
\begin{eqnarray} \label{scGeqs}
\scG_{1-2j}^{(\al)}(x)-\intinf \! dy \, G_\al(x-y)\, \mm(y) \, \scG_{1-2j}^{(\al)}(y)&=&e^{-(2j-1)\,x}, \\
\frac{\scG_{1-2j,m}^{(\al)}(x)}{\mm(x)}-\intinf \! dy \, G_\al(x-y)\,  \scG_{1-2j,m}^{(\al)}(y)&=&e^{-(2j-1)\, x},
\label{scGeqsm}
\end{eqnarray}
where $G_\al(x)$ is given in (\ref{Galpha}).
In the papers \cite{Jimbo:2010jv,Hegedus:2019rju}, $\om_{2k-1,1-2j}(\al)$ is originally defined in terms of $\scG_{1-2j}^{(\al)}(x).$
Nevertheless, in our subsequent computations and considerations, the formulation 
in terms of $\scG_{1-2j,m}^{(\al)}(x)$ 
proves to be useful. The reason for this is that the integral operator in (\ref{scGeqsm}) has useful 
symmetry properties under transposition. If (\ref{scGeqsm}) is written in operator form:
\begin{equation} \label{Opeq}
\begin{split}
\intinf \! dy \, {\cal M}_\al(x,y) \, \scG_{1-2j,m}^{(\al)}(y)=e^{-(2j-1)\,x}, 
\end{split}
\end{equation}
where
\begin{equation} \label{cMal}
\begin{split}
{\cal M}_\al(x,y)=\frac{\delta(x-y)}{\mm(x)}-G_\al(x-y),
\end{split}
\end{equation}
denotes the kernel of the corresponding integral operator\footnote{Here, $\delta$ denotes the Dirac delta.}, 
then the transposed kernel differs from the original one only by the replacement $\al \to -\al$: 
\begin{equation} \label{cMalTR}
\begin{split}
{\cal M}_\al(y,x)={\cal M}_{-\al}(x,y).
\end{split}
\end{equation}
This means that, at $\al=0$, it is a symmetric operator. These transposition and symmetry properties will 
play an important role in our subsequent considerations. Here, an important remark is in order: what we now call 
$G_{-\al}(x)$ is defined by (\ref{Galpha}) with the replacement $\al \to -\al$ for $\al>0.$ This is not equal to $G_\al(x)$ 
analytically continued to the negative value $-\al.$


Before entering the details of the derivation, let us emphasize that the integration measure $\mm(x)$ is determined by the solution of the finite-volume NLIE (\ref{DDV}). Denoting this solution by $Z(x)$, the measure is defined by (\ref{m}) as
\begin{equation} \label{mZx}
\begin{split}
\mm(x)\equiv\mm[Z(x)]=\scL_+[Z(x+i\, 0)]+\scL_-[Z(x-i\, 0)], \quad \mbox{with} \quad 
\scL_\pm[Z(x)]=\frac{e^{\pm \, i\,Z(x)}}{1+e^{\pm \, i\,Z(x)}}.
\end{split}
\end{equation}
Thus, the measure can naturally be viewed as a functional of an arbitrary function. From this perspective, several measure functions associated with particular solutions become relevant for the UV analysis:
\begin{itemize}
\item the exact finite-volume measure
\[
\mm(x)=\mm[Z(x)],
\]

\item the kink measures associated with the kink solutions $Z_\pm(x)$ defined in (\ref{Zpmdef}),
\[
\mm_\pm(x)=\mm[Z_\pm(x)],
\]

\item and the measure corresponding to the asymptotic solution (\ref{Zasdef}),
\[
\mm_{as}(x)=\mm[Z_{as}(x)].
\]
\end{itemize}

The measures $\mm(x)$ and $\mm_{as}(x)$ both depend on $\ell$ and exhibit double-exponential decay as $x\to\pm\infty$. Consequently, integrals involving these functions remain convergent even when multiplied by exponentially growing factors.

By contrast, the kink measures $\mm_\pm(x)$ are $\ell$-independent and decay double-exponentially only in one direction: $\mm_+(x)$ as $x\to+\infty$ and $\mm_-(x)$ as $x\to-\infty$. In the opposite directions, corresponding to the plateau regimes, they approach a constant determined by the plateau value $z_0$ given in (\ref{z0value}).

As a consequence, the integrals arising in the UV analysis may become divergent on the plateau side, and therefore require a more careful treatment in the $\ell\to0$ limit.

With the help of these measures, one can define the important kink linear problems, whose solutions 
enter the final formulas of the UV expansion. These are as follows\footnote{These equations agree with those 
defined in (\ref{kinkmalpha}) and (\ref{kinkpalpha}).}: 
\begin{equation} \label{scGkinkpmj}
\begin{split}
\scG_{\pm j}^{(\al),(\pm)}(x)-\intinf \! dy \, G_\al(x-y)\, \mm_\pm(y) \, \scG_{\pm j}^{(\al),(\pm)}(x)=e^{\pm j\, x}, 
\qquad j=1,3,5,7,...
\end{split}
\end{equation}
where, in the notation for the solution function $\scG_{\pm j}^{(\al),(\pm)}(x),$ the superscript 
refers to the kink regime specified by the measure $\mm_\pm(x),$ while the subscript $\pm j$ refers to the 
source term $e^{\pm j\, x}.$

The UV expansion of $\tom_{2k-1,1-2j}(\al)$ in the $\ell\to0$ limit is based on the assumption that the asymptotic solution provides an accurate approximation to the exact finite-volume solution in the UV regime. More precisely, we assume that the leading UV behaviour is entirely encoded in the asymptotic solution, while the difference between the exact and asymptotic solutions yields only subleading corrections.

Accordingly, the expansion proceeds in two steps. First, we expand around the asymptotic solution, i.e. in powers of
\[
\delta Z(x)=Z(x)-Z_{as}(x).
\]
It is assumed that successive terms in this expansion become increasingly suppressed in the small-volume limit. Nevertheless, each term still possesses a nontrivial $\ell$ dependence that must be extracted systematically.

Therefore, in the second step, these terms are themselves expanded around the leading asymptotic solution defined later in (\ref{asscG0}) and (\ref{scGlas}), in order to make their explicit $\ell$ dependence manifest.

In the following, we formulate these steps in more precise mathematical terms.

The corrections to the leading asymptotic term originate from the corrections to the measure:
$\mm(x)=\mm_{as}(x)+\delta \mm(x),$  which follows from the deviation between  
$Z(x)$ and $Z_{as}(x),$ which becomes small in the UV limit. 


We start with the definition of the asymptotic solutions of equations (\ref{scGeqs}) and (\ref{scGeqsm}). 
They are defined as the solutions of the corresponding asymptotic equations:
\begin{eqnarray} \label{scGeqsAS}
\scG_{1-2j}^{(\al),(as)}(x)&-&\intinf \! dy \, G_\al(x-y)\, \mm_{as}(y) \, \scG_{1-2j}^{(\al),(as)}(y)=e^{-(2j-1)x}, \\
\frac{\scG_{1-2j,m}^{(\al),(as)}(x)}{\mm_{as}(x)}&-&\intinf \! dy \, G_\al(x-y)\,  \scG_{1-2j,m}^{(\al),(as)}(y)=e^{-(2j-1)x}, 
\label{scGeqsmAS} \\
\scG_{1-2j,m}^{(\al),(as)}(x)&=& \mm_{as}(x) \, \scG_{1-2j}^{(\al),(as)}(x).
\end{eqnarray}
The functions $\scG_{1-2j}^{(\al),(as)}(x)$ and $\scG_{1-2j,m}^{(\al),(as)}(x)$ are called the asymptotic solutions
of the linear problems (\ref{scGeqs}) and (\ref{scGeqsm}), respectively. They contain the leading 
$\ell \to 0$ behaviour of the problem and thus form the background around which 
the original problem is expanded in the UV limit. Nevertheless, it is obvious that these functions possess 
nontrivial $\ell$ dependence, which requires further careful treatment in order to obtain the UV series expansion in $\ell$. 

In principle, one should solve (\ref{scGeqsAS}) or (\ref{scGeqsmAS}) to determine the asymptotic solution. 
Nevertheless, these asymptotic equations are still very complicated and cannot be solved exactly. 
The only thing we can do is to expand their solutions around the ``leading asymptotic solutions'', which are 
known functions in principle. They capture the leading small-$\ell$ behaviour of these solutions.

They can be obtained by a simple line of reasoning \cite{enjhep}. The source term in both (\ref{scGeqsAS}) and (\ref{scGeqsmAS}) is $e^{-(2j-1)x}$. Its dominant contribution arises from the $(-\infty,0)$ regime, where, at leading order, one may replace $\mm_{as}(x)$ with $\mm_-(x+\log\tfrac{2}{\ell})$.

Performing this replacement in (\ref{scGeqsAS}), making the change of variables $x \to x - \ltl,$ and 
comparing the result to the kink equation (\ref{scGkinkpmj}), one can conclude that the leading 
asymptotic solutions of (\ref{scGeqsAS}) and (\ref{scGeqsmAS}) can be expressed in terms of 
the kink solutions of (\ref{scGkinkpmj}) as follows: 
\begin{equation} \label{asscG}
\begin{split}
\scG_{1-2j}^{(\al),(as)}(x)&\simeq \left(\frac{2}{\ell}\right)^{2j-1} \, \scG_{1-2j}^{(\al),(-)}(x+\log \tfrac{2}{\ell}),  \\
\scG_{1-2j,m}^{(\al),(as)}(x)&\simeq \left(\frac{2}{\ell}\right)^{2j-1} \mm_{as}(x) \, \scG_{1-2j}^{(\al),(-)}(x+\log \tfrac{2}{\ell}).
\end{split}
\end{equation}
Since we will use these functions extensively, we introduce a notation for them by placing a $0$ subscript 
on the asymptotic solutions. Namely:
\begin{equation} \label{asscG0}
\begin{split}
\scG_{1-2j,0}^{(\al),(as)}(x)&= \left(\frac{2}{\ell}\right)^{2j-1} \, \scG_{1-2j}^{(\al),(-)}(x+\log \tfrac{2}{\ell}),  \\
\scG_{1-2j,m,0}^{(\al),(as)}(x)&= \left(\frac{2}{\ell}\right)^{2j-1} \mm_{as}(x) \, \scG_{1-2j}^{(\al),(-)}(x+\log \tfrac{2}{\ell}).
\end{split}
\end{equation}
We are now in a position to formulate the expansion of the original linear problem (\ref{scGeqsm}). 
As a first step, we expand it around the leading asymptotic solution $\scG_{1-2j,m,0}^{(\al),(as)}(x).$ 
We identify two basic types of corrections to this background: 
\begin{equation} \label{dscG}
\begin{split}
\scG_{1-2j,m}^{(\al)}(x)=\scG_{1-2j,m,0}^{(\al),(as)}(x)+\delta \scG_{1-2j,m}^{(\al),(as)}(x)+\delta \scG_{1-2j,m}^{(\al),\delta m}(x).
\end{split}
\end{equation}
The first correction, $\delta \scG_{1-2j,m}^{(\al),(as)}(x),$ originates purely from the asymptotic equation 
(\ref{scGeqsmAS}), while the second one, $\de \scG_{1-2j,m}^{(\al),\delta m}(x),$ originates from the corrections to the measure in the asymptotic equation. 

In \cite{enjhep} the linear problems satisfied by the functions appearing on the right-hand side of (\ref{dscG}) 
were derived. We now present them below. 
The leading asymptotic solution satisfies the following equation:
\begin{equation} \label{scGmas0}
\begin{split}
\frac{\scG_{1-2j,m,0}^{(\al),(as)}(x)}{\mm_{as}(x)}-\intinf \! dy \, G_\al(x-y)\, 
\frac{\mm_-(y+\log\tfrac{2}{\ell})}{\mm_{as}(y)} \, \scG_{1-2j,m,0}^{(\al),(as)}(y)=e^{-(2j-1)\, x}.
\end{split}
\end{equation}
The asymptotic correction,
$\delta \scG_{1-2j,m}^{(\al),(as)}(x)=\scG_{1-2j,m}^{(\al),(as)}(x)- \scG_{1-2j,m,0}^{(\al),(as)}(x),$
is the solution of the following linear problem:
\begin{equation} \label{dscGmas}
\begin{split}
\frac{\de \scG_{1-2j,m}^{(\al),(as)}(x)}{\mm_{as}(x)}-\intinf \! dy \, G_\al(x-y)\,  
\de \scG_{1-2j,m}^{(\al),(as)}(y)=\de RH^{(as)}_{1-2j}(x), \\
\de RH^{(as)}_{1-2j}(x)=\left(\frac{2}{\ell}\right)^{2j-1} \intinf \! dy \, G_\al(x-y) \, \scG_{1-2j}^{(\al),(-)}(y+\log\tfrac{2}{\ell}) \, 
\left[ \mm_{as}(y)-\mm_-(y+\log\tfrac{2}{\ell}) \right].
\end{split}
\end{equation}
Finally, $\de \scG_{1-2j,m}^{(\al),\delta m}(x)$ satisfies a linear equation whose source term is proportional 
to the correction of the measure $\delta \mm(x)=\mm(x)-\mm_{as}(x):$
\begin{equation} \label{dscGdm}
\begin{split}
\frac{\de \scG_{1-2j,m}^{(\al),\de m}(x)}{\mm_{as}(x)+\delta \mm(x)}-\intinf \! dy \, G_\al(x-y)\,  
\de \scG_{1-2j,m}^{(\al),\de m}(y)=\frac{\de \mm(x)}{\mm_{as}(x)\, (\mm_{as}(x)+\delta \mm(x))} \, \scG_{1-2j,m}^{(\al),(as)}(x).
\end{split}
\end{equation}
This function is proportional to the corrections to the asymptotic solution (\ref{Zasdef}); 
therefore, we expect it not to contribute to the leading UV behaviour. 

Based on the representation (\ref{dscG}), in the UV limit $\tom_{2k-1,1-2j}(\al)$ can be written as the sum of three terms:
\begin{equation} \label{A20}
\begin{split}
\tom_{2k-1,1-2j}(\al)\! &=\tom_{0,kj}^{(as)}(\al)+\de \tom^{(as)}_{kj}(\al)+
\de \tom^{\de m}_{kj}(\al), \\
\tom_{0,kj}^{(as)}(\al)\! &=
\!\!\!\!\intinf \!\! dx \, e^{(2k-1)x} \, \scG_{1-2j,m,0}^{(\al),(as)}(x), 
\qquad \de \tom^{(as)}_{kj}(\al)\!=\!\!\!\!\intinf \!\! dx \, e^{(2k-1)x} \,\de \scG_{1-2j,m}^{(\al),(as)}(x), \\
\de \tom^{\de m}_{kj}(\al)\! &= \!\!\!\!\intinf \!\! dx \, e^{(2k-1)x} \, \de \scG_{1-2j,m}^{(\al),\de m}(x). 
\end{split}
\end{equation}
The contribution $\de \tom^{\de m}_{kj}(\al)$ does not contribute to the leading UV formula for vacuum  expectation values; thus, 
we focus only on the first two terms in (\ref{A20}) in order to capture the terms constituting 
$\tom_{2k-1,1-2j}^{eff}(\alpha)$. As a first step, we rewrite them in forms appropriate for the UV expansion. 

The first term does not require any special treatment. 
It is governed solely by the leading asymptotic solution:
\begin{equation} \label{tomas0}
\begin{split}
\tom_{0,kj}^{(as)}(\al)\! =\!\!\!\! \intinf \!\! dx \, e^{(2k-1)x} \, \scG_{1-2j,m,0}^{(\al),(as)}(x)\!=\left(\frac{2}{\ell}\right)^{2k-1}
\!\! \intinf \!\! dx \, e^{(2k-1)\,x} \, \mm_{as}(x) \, \scG_{1-2j}^{(\al),(-)}(x+\log\tfrac{2}{\ell}).
\end{split}
\end{equation}
The next term is:
\begin{equation} \label{dtomasdef}
\begin{split}
\de \tom^{(as)}_{kj}(\al)=\!\!\!\! \intinf \!\! dx \, e^{(2k-1)x} \, \de \scG_{1-2j,m}^{(\al),(as)}(x).
\end{split}
\end{equation}
To rewrite it in a form appropriate for the UV expansion, we represent the solution of (\ref{dscGmas}) 
formally using the asymptotic counterpart of the linear operator (\ref{cMal}) of the problem\footnote{
By the asymptotic counterpart we mean the operator
${\cal M}^{(as)}_\al(x,y)=\frac{\de(x-y)}{\mm_{as}(x)}-G_\al(x-y),$
which possesses the same transposition properties as  
${\cal M}_\al(x,y).$ See (\ref{cMalTR}).}:
\begin{equation} \label{dscGmasformal}
\begin{split}
\de \scG_{1-2j,m}^{(\al),(as)}(x)=\!\! \intinf \! dy \,{\cal M}_\al^{(as),-1}(x,y) \, \de RH^{(as)}_{1-2j}(y).
\end{split}
\end{equation}
Inserting (\ref{dscGmasformal}) into (\ref{dtomasdef}) and exploiting the transposition property (\ref{cMalTR}),  
one obtains the following formula:
\begin{equation} \label{dtomasfin}
\begin{split}
\de \tom^{(as)}_{kj}\!(\al)\!=\!\! \left(\frac{2}{\ell}\right)^{2j-1} \!\!\!\! \intinf \!\!\!\! dx \,  \scG_{1-2j}^{(\al),(-)}(x+\log\tfrac{2}{\ell}) \!\! 
\left[ \mm_{as}(x)\!-\!\mm_-(x\!+\!\log\tfrac{2}{\ell}) \! \right] \!\!\!
\left[ \!\frac{\scG_{2k-1,m}^{(-\al),(as)}(x)}{\mm_{as}(x)}\!-\!e^{(2k-1)x} \! \right]\!\!,
\end{split}
\end{equation}
where the action of the inverse transpose of ${\cal M}_\al^{(as)}(x,y)$ on $e^{(2k-1)x}$ is defined by the function 
\begin{equation} \label{dscGp1maldef}
\begin{split}
\scG_{2k-1,m}^{(-\al),(as)}(x)=\intinf \! dy \, {\cal M}_\al^{(as),-1}(y,x) \, e^{(2k-1)y}.
\end{split}
\end{equation}
As a consequence, it satisfies a linear equation, which we also exploited when deriving (\ref{dtomasfin}):
\begin{equation} \label{dscGp1mal}
\begin{split}
\intinf dy \, G_{-\al}(x-y) \scG_{2k-1,m}^{(-\al),(as)}(y)=
\frac{\scG_{2k-1,m}^{(-\al),(as)}(x)}{\mm_{as}(x)}-e^{(2k-1)x}.
\end{split}
\end{equation}
To treat (\ref{dscGp1mal}) in the UV limit, we again define the corresponding leading asymptotic solution:
\begin{equation} \label{scGlas}
\begin{split}
\scG_{2k-1,m,0}^{(-\al),(as)}(x)=\left(\frac{2}{\ell}\right)^{2k-1} \!\!\! \mm_{as}(x) \, \scG_{2k-1}^{(-\al),(+)}(x-\log\tfrac{2}{\ell}),
\end{split}
\end{equation}
with $\scG_{2k-1}^{(-\al),(+)}(x)$ being the solution of (\ref{kinkpalpha}).
This definition follows from a line of reasoning similar to that which led to (\ref{asscG}).

The proper solution of (\ref{dscGp1mal}) is sought as an expansion around this background:
\begin{equation} \label{scGpm0def}
\begin{split}
\scG_{2k-1,m}^{(-\al),(as)}(x)=\scG_{2k-1,m,0}^{(-\al),(as)}(x)+\de \scG_{2k-1,m}^{(-\al),(as)}(x).
\end{split}
\end{equation}
With the help of (\ref{dscGp1mal}), one can derive the linear equations satisfied by  
the leading-order background $\scG_{2k-1,m,0}^{(-\al),(as)}(x)$ 
and the correction $\de \scG_{2k-1,m}^{(-\al),(as)}(x).$ 

The linear equation for the leading-order background takes the form:
\begin{equation} \label{scGpas0}
\begin{split}
\frac{\scG_{2k-1,m,0}^{(-\al),(as)}(x)}{\mm_{as}(x)}-\intinf \! dy \, G_{-\al}(x-y)\, 
\frac{\mm_+(y-\log\tfrac{2}{\ell})}{\mm_{as}(y)} \, \scG_{2k-1,m,0}^{(-\al),(as)}(y)=e^{(2k-1)x}.
\end{split}
\end{equation}
The asymptotic correction,
$\delta \scG_{2k-1,m}^{(-\al),(as)}(x)=\scG_{2k-1,m}^{(-\al),(as)}(x)- \scG_{2k-1,m,0}^{(-\al),(as)}(x),$ 
is the solution of the following linear problem:
\begin{equation} \label{dscGpas}
\begin{split}
\frac{\de \scG_{2k-1,m}^{(-\al),(as)}(x)}{\mm_{as}(x)}-\intinf \! dy \, G_{-\al}(x-y)\,  
\de \scG_{2k-1,m}^{(-\al),(as)}(y)=\de RH^{(-\al)}_{2k-1}(x), \\
\de RH^{(-\al)}_{2k-1}(x)=\left(\frac{2}{\ell}\right)^{2k-1} \!\!\! \intinf \! dy \, G_{-\al}(x-y) \, \scG_{2k-1}^{(-\al),(+)}(y-\log\tfrac{2}{\ell}) \, 
\left[ \mm_{as}(y)-\mm_+(y-\log\tfrac{2}{\ell}) \right].
\end{split}
\end{equation}
The formulas (\ref{dscGp1mal})--(\ref{dscGpas}) will play an important role in the explicit 
computation of the coefficients in the $\ell \to 0$ series representation of $\de \tom^{(as)}_{kj}(\al).$
 
The formulas (\ref{tomas0}) and (\ref{dtomasfin}) constitute the 
starting point for the subsequent steps of determining the coefficients of the UV expansion of 
$\tom_{2k-1,1-2j}^{eff}(\al).$  The small-volume expansion of these formulas makes it possible 
to identify, among the many different contributions, the ones that build up $\tom_{2k-1,1-2j}^{eff}(\al).$

\subsection{The basic strategy for computing the coefficients of the UV series }

The three contributions appearing in (\ref{A20}) exhibit a nontrivial dependence on $\ell$, which must be extracted in a systematic way. In this subsection, following \cite{enjhep}, we outline the general strategy for computing the $\ell \to 0$ expansion of $\tom_{2k-1,1-2j}(\alpha)$. At each order of the small-volume expansion, each term in (\ref{A20}) decomposes into a sum of several contributions. Each of these is given by integral expressions involving products of left- and right-mover combinations of $\ell$-independent kink functions.\footnote{Here, an $\ell$-independent function $f$ is called a left-mover (right-mover)
if it appears with argument $x+\log\tfrac{2}{\ell}$ $\left(x-\log\tfrac{2}{\ell}\right).$} 
Consequently, all $\ell$-dependence enters exclusively through the shifted arguments of these functions.

To disentangle this structure, we proceed in the following way. Each $\intinf$ integral is split into two pieces:
\[
\intinf = \int\limits_{-\infty}^0 + \int\limits_0^\infty.
\]
Consider first the integral over $(0,\infty).$ After the change of variables
\[
x \to x+\log\tfrac{2}{\ell},
\]
the integration domain becomes $\int\limits_{-\log\tfrac{2}{\ell}}^\infty.$ This transformation has two important consequences. First, the arguments of the right-mover ``+'' kink functions become independent of $\ell.$ Second, the ``$-$'' kink functions are shifted into their plateau regime, where they admit a systematic small-$\ell$ expansion. 

Expanding the ``$-$'' kink functions generates explicit powers of $\ell,$ while the remaining integrand becomes $\ell$ independent. As a consequence, the $\ell$ dependence of the integral originates from two sources only:
\begin{itemize}
\item explicit powers of $\ell$ produced by the plateau expansion,
\item the $\ell$ dependence of the lower integration bound.
\end{itemize}

Since the integrand itself is $\ell$ independent after the transformation, the contribution coming from the lower bound can be extracted directly from the asymptotic expansion of the integrand in the $x\to -\infty$ regime. According to our assumptions, this expansion is known in principle, since it corresponds precisely to the plateau behaviour of the kink functions.

Using this procedure, one can systematically derive the small-$\ell$ expansion of any of the integrals appearing in (\ref{A20}). 

The treatment of the integrals over $(-\infty,0)$ is completely analogous. Therefore, the UV expansion of each of the three terms in (\ref{A20}) can be obtained within the same framework. 

Now we apply the method  in detail to the quantities 
stemming from the asymptotic solution.{\footnote{Namely, to $\tom_{0,kj}^{(as)}(\al)$ and $\de \tom_{kj}^{(as)}(\al).$}}

\subsection{The computation of the $\ell \to 0$ series of  $\tom_{0,kj}^{(as)}(\al)$}

 We proceed as described in the previous subsection.

Starting from (\ref{tomas0}), splitting the integral into two parts and performing the appropriate shifts, $\tom_{0,kj}^{(as)}(\al)$ can be written as the sum of two contributions:
\begin{equation} \label{tomas0sum}
\begin{split}
\tom_{0,kj}^{(as)}(\al)=\tom_{0,kj}^{(as),+}(\al)+\tom_{0,kj}^{(as),-}(\al),
\end{split}
\end{equation}
where 
\begin{eqnarray}\label{tomas0p}
\tom_{0,kj}^{(as),+}(\al)&=& \left(\frac{2}{\ell}\right)^{2(j+k-1)} \!\!\!\!\!  \int\limits_{-\log\tfrac{2}{\ell}}^\infty \!\!\! dx \, e^{(2k-1)x} \, \mm_{as}(x+\log\tfrac{2}{\ell}) \,
\scG_{1-2j}^{(\al),(-)}(x+2 \log\tfrac{2}{\ell}),\\
\tom_{0,kj}^{(as),-}(\al)&=&\int\limits_{-\infty}^{\log\tfrac{2}{\ell}} \! dx \, e^{(2k-1)\,x} \, \mm_{as}(x-\log\tfrac{2}{\ell}) \,
\scG_{1-2j}^{(\al),(-)}(x).
\label{tomas0m}
\end{eqnarray}

We first consider $\tom_{0,kj}^{(as),+}(\al).$
Using the plateau-regime expansions (\ref{kinkplat}) and (\ref{scGmasy}), one obtains the following small-$\ell$ series representations in the range $-\ltl<x<\infty$:
\begin{equation} \label{smallell1}
\begin{split}
\scG_{1-2j}^{(\al),(-)}(x+2 \, \ltl) &\simeq \sum\limits_{n=1}^\infty  \left(\frac{2}{\ell}\right)^{\tfrac{2 \al-4 \,n}{p+1}} \, c_n^{(-(2j-1),\al)} \, e^{\tfrac{\al-2 \,n}{p+1}x}, \\
Z_-(x+2 \, \ltl)-z_0 &\simeq \sum\limits_{n=1}^\infty 
\left(\frac{\ell}{2}\right)^{\tfrac{4 n}{p+1}} \, c_n \,  e^{-\frac{2\, n \, x}{p+1}}, \\
\mm_{as}(x+\ltl) &= \mm[Z_+(x)+Z_-(x+ 2 \, \ltl)-z_0]\simeq\mm_+(x)\!+\!\!\sum\limits_{n=1}^\infty 
\left(\frac{\ell}{2}\right)^{\tfrac{4 n}{p+1}} \, \mm^+_{as,n}(x),
\end{split}
\end{equation}
where the coefficients $\mm^+_{as,n}(x)$ can be expressed in terms of $c_n$ and derivatives of the measure,
$\frac{d^s}{dZ^s}\mm[Z(x)]\big|_{Z=Z_+},$ although their explicit form is not needed for our purposes.  

Substituting (\ref{smallell1}) into (\ref{tomas0p}), one obtains the structure
\begin{equation} \label{tomas0pfin0}
\begin{split}
\tom_{0,kj}^{(as),+}(\al)\!\simeq \! \left(\frac{2}{\ell}\right)^{\!\! 2(j+k-1)+\tfrac{2 \al}{p+1}} \!\!
\sum\limits_{n=1}^\infty \left(\frac{\ell}{2}\right)^{\tfrac{4 \,n}{p+1}}  \!\!\!
\left\{  c_n^{-(2j-1),\al)} \!\!\!\!\! \intli \!\!\!\! dx \, e^{(2k-1)x} 
\, \mm_+(x) \, e^{\tfrac{\al-2 \,n }{p+1} x}\!+\!\dots \!\!
 \right\}\!,
\end{split}
\end{equation}
where the dots denote terms proportional to $c_s^{(-(2j-1),\al)}$ with $s\leq n-1.$
The explicitly displayed contribution is precisely the one entering $\tom_{2k-1,1-2j}^{eff}(\al).$
Comparing (\ref{tomas0pfin0}) with (\ref{UVexpansion}), it follows that one of the contributions
to the $n$th term of $\tom_{2k-1,1-2j}(\al)$ originating from $\tom_{0,kj}^{(as),+}(\al)$ is
$c_n^{(-(2j-1),\al)} \, I_{n,0}^{(k)},$ with $I_{n,0}^{(k)}$ defined in (\ref{InkCP}).
Based on our numerical observations, the remaining terms do not contribute to the
vacuum expectation values at leading order in the UV limit. Therefore,
$c_n^{(-(2j-1),\al)} \, I_{n,0}^{(k)}$ is the only contribution of
$\tom_{0,kj}^{(as),+}(\al)$ to $\tom_{2k-1,1-2j|n}^{eff}(\al)$.

As argued in \cite{enjhep} the contributions arising from $\tom_{0,kj}^{(as),-}(\al)$ are not consistent with the field-theoretic expectations for the small-$\ell$ behaviour, and are therefore expected to cancel against contributions from other sources.

\subsection{The computation of the $\ell \to 0$ series of  $\delta \tom^{(as)}_{kj}(\al)$}

Our starting point is (\ref{dtomasfin}). We first decompose it into two contributions:
\begin{equation} \label{dtomassum}
\begin{split}
\delta \tom^{(as)}_{kj}(\al)=\delta \tom_{0,kj}^{(as)}(\al)+\delta \tom_{\de,kj}^{(as)}(\al).
\end{split}
\end{equation}

One of these terms involves only the leading-order asymptotic solutions:
\begin{equation} \label{dtomas00}
\begin{split}
\de \tom_{0,kj}^{(as)}(\al)&=\left(\frac{2}{\ell}\right)^{\!\! 2j-1} \! \intinf \! dx \, \scG_{1-2j}^{(\al),(-)}(x+\log\tfrac{2}{\ell}) \, 
\left[ \mm_{as}(x)-\mm_-(x+\log\tfrac{2}{\ell}) \right] \,\times \\
&\times \left[\left(\frac{2}{\ell}\right)^{\!\! 2k-1} \, \scG_{2k-1}^{(-\al),(+)}(x-\ltl)-e^{(2k-1)x} \right].
\end{split}
\end{equation}
The other term is proportional to the asymptotic correction:
\begin{equation} \label{dtomasdG}
\begin{split}
\de \tom_{\de,kj}^{(as)}(\al)=\left(\frac{2}{\ell}\right)^{2j-1} \! \intinf \! dx \, \scG_{1-2j}^{(\al),(-)}(x+\log\tfrac{2}{\ell}) \, 
\left[ \mm_{as}(x)-\mm_-(x+\log\tfrac{2}{\ell}) \right] \,
 \frac{\de \scG_{2k-1,m}^{(-\al),(as)}(x)}{\mm_{as}(x)}.
\end{split}
\end{equation}
Our numerical computations indicate that (\ref{dtomasdG}) does not contribute to the leading-order UV term of the vacuum expectation values. We therefore do not analyze its UV expansion analytically.
Thus, to obtain the leading-order contribution, only (\ref{dtomas00}) needs to be considered. As explained in the previous section, we decompose it further into two integrals:
\begin{equation} \label{dtomas0}
\begin{split}
\de \tom_{0,kj}^{(as)}(\al)=\de \tom_{0,kj}^{(as),+}(\al)+\de \tom_{0,kj}^{(as),-}(\al),
\end{split}
\end{equation}
where
\begin{equation} \label{dtomas0p}
\begin{split}
\de \tom_{0,kj}^{(as),+}(\al)&\!=\!\!\left(\frac{2}{\ell}\right)^{\!\! 2(j+k-1)} \!\!\!\!\!\! \intli \!\!\!\!\! dx \,
 \scG_{1-2j}^{(\al),(-)}(x+2\, \log\tfrac{2}{\ell}) \, 
\left[ \mm_{as}(x+\ltl)-\mm_-(x+2 \, \log\tfrac{2}{\ell}) \right]  \\
&\times \left[ \scG_{2k-1}^{(-\al),(+)}(x)-e^{(2k-1)x} \right],
\end{split}
\end{equation}
and
\begin{equation} \label{dtomas0m}
\begin{split}
\de \tom_{0,kj}^{(as),-}(\al)\!&=\!\left(\frac{2}{\ell}\right)^{2j-1} \!\!\! \intil \!\! dx \,
 \scG_{1-2j}^{(\al),(-)}(x)  \left[ \mm_{as}(x-\ltl)\!-\!\mm_-(x) \right] \!   \\
 & \times \, \left[ \left(\frac{2}{\ell}\right)^{2k-1} \, \scG_{2k-1}^{(-\al),(+)}(x-2 \, \ltl)\!-\!\left(\frac{\ell}{2}\right)^{2k-1} e^{(2k-1)x} \right].
\end{split}
\end{equation}

We begin with $\de \tom_{0,kj}^{(as),+}(\al).$ Using the expansions collected in (\ref{smallell1}), one obtains:
\begin{equation} \label{dtomas0pres}
\begin{split}
\de \tom_{0,kj}^{(as),+}(\al)&=\left(\frac{2}{\ell}\right)^{2(j+k-1)+\tfrac{2 \al}{p+1}} 
\sum\limits_{n=1}^\infty \,\left(\frac{\ell}{2}\right)^{\tfrac{4 n}{p+1}} 
\bigg\{
\, c_n^{(-(2j-1),\al)}\! \\ 
\times & \int\limits_{-\ltl}^\infty  \! dx \,\, e^{\tfrac{\al-2\,n}{p+1}x} \, 
\left( \mm_{+}(x)-\mm[z_0] \right)\, \left( \scG_{2k-1}^{(-\al),(+)}(x)-e^{(2k-1)x} \right)+\dots
\bigg\}.
\end{split}
\end{equation}
where the dots denote terms proportional to $c_s^{(-(2j-1),\al)}$ with $s\leq n-1.$
The explicitly displayed contribution is the one entering $\tom_{2k-1,1-2j}^{eff}(\al).$
Comparing (\ref{dtomas0pres}) with (\ref{UVexpansion}), it follows that one of the contributions
to the $n$th term of $\tom_{2k-1,1-2j}(\al)$ originating from $\de \tom_{0,kj}^{(as),+}(\al)$ is
$c_n^{(-(2j-1),\al)} \, I_{n,1}^{(k)},$ with $I_{n,1}^{(k)}$ defined in (\ref{InkCP}).

Based on our numerical studies, the remaining terms, as well as the contributions coming from
$\de \tom_{0,kj}^{(as),-}(\al)$ defined in (\ref{dtomas0m}), do not contribute to the vacuum expectation values
at leading order in the UV limit, and therefore do not contribute to
$\tom_{2k-1,1-2j}^{eff}(\al).$

Just to summarize, in this appendix we have analyzed a general term in the systematic UV expansion (\ref{UVexpansion}) of $\tom_{2k-1,1-2j}(\alpha)$. We focused on those contributions that are fully determined by the asymptotic solution (\ref{Zasdef}). We showed that, at order $n$, among the large number of terms, one can naturally identify a specific contribution which, according to our numerical evidence, is the only one that contributes at leading order to the vacuum expectation values in the UV limit. By definition, these leading contributions form the quantity $\tom_{2k-1,1-2j}^{eff}(\alpha)$.

We emphasize that we do not derive the formula (\ref{tomeffn}) for $\tom_{2k-1,1-2j}^{eff}(\alpha)$. What we establish is only that the terms appearing in (\ref{tomeffn}) correspond to a single distinguished contribution within the full set of terms that make up the $n$th - order coefficient in the UV expansion of $\tom_{2k-1,1-2j}(\alpha)$, namely the contribution  that determines the leading UV behaviour of vacuum expectation values.

We are not able to prove analytically that the remaining contributions cancel in the leading-order 
UV expression for vacuum expectation values. This cancellation is supported solely by high-precision numerical evidence, with agreement up to at least 19-digit precision.


\section{The large argument expansion of the kernel $G_\al(x)$} \label{appB}

The knowledge of the large argument series expansion of the kernel $G_\al(x)$ (\ref{Galpha}), is useful for two reasons. 
First, it offers an alternative useful series representation of the kernel for its numerical computation 
in a regime, where the accurate numerical evaluation of the original Fourier-representation (\ref{Galpha}) 
becomes challenging due to the highly oscillating nature of the integrand. 

Second, such a series representation of $G_\al(x),$ allows one to compute the qualitative large argument 
behaviours of the kink functions from the linear equations they satisfy. Namely, the second series 
representations in (\ref{scGmasy}) and (\ref{scGpasy}) can be derived from the equations 
(\ref{kinkmalpha}) and (\ref{kinkpalpha}) respectively, 
with the help of the large argument series of $G_\al(x).$

The large-argument asymptotic expansion of $G_\al(x)$ can be derived from its Fourier-integral representation (\ref{Galpha}) by means of the residue theorem.
We present the corresponding series expansions at $\pm\infty$, which take the form
\begin{equation} \label{Gsoral}
\begin{split}
G_\al(x)&\stackrel{x\to +\infty}{=}\sum\limits_{k=0}^\infty \hat{C}_k^{(\al)} \, e^{-(1+2k) x} +
\sum\limits_{k=1}^\infty \tilde{C}_k^{(\al)} \, e^{-\tfrac{2k-\al}{p} x},  
\qquad 0\leq \al <2, \\ 
G_\al(x)&\stackrel{x\to -\infty}{=}-\sum\limits_{k=0}^\infty \hat{C}_{-1-k}^{(\al)} \, e^{(1+2k) x} -
\sum\limits_{k=0}^\infty \tilde{C}_{-k}^{(\al)} \, e^{\tfrac{2k+\al}{p} x},  
\qquad 0\leq \al <2, \\
\mbox{with} \qquad
\hat{C}_k^{(\al)}&=-\frac{1}{\pi} \cot\left[ (1+2 k) \tfrac{\pi \, p}{2}+\al \right],  \qquad
\tilde{C}_k^{(\al)}=\frac{1}{p \, \pi} \tan\left[ \tfrac{\pi}{2 \, p} (2k-\al) \right].
\end{split}
\end{equation}
We note that throughout this paper we use the notation
$G_{-\al}(x)=G_\al(-x)$, which is obtained by the formal
substitution $\al\to-\al$ in the Fourier representation
(\ref{Galpha}).
This should not be confused with the analytic continuation
of the function $G_\al(x)$ from positive to negative values
of $\al$, since the two procedures do not yield the same
result.


\section{Numerical validation of the main result} \label{appC}

In this appendix, we present explicit numerical data supporting the main result
(\ref{calCres}) of the paper.

For each parameter set considered in this appendix, we provide the set of
coefficients $c_n^{(-(2j-1),\al)}$ appearing in the plateau expansion
(\ref{scGmasy}), the numerical values of the integrals (\ref{InkCP}), and
finally a numerical comparison between the leading UV coefficient
(\ref{calCres}) and the analytical CFT prediction (\ref{simVEV1}). 

As a first example, we consider the parameter choice $(\al,p,\al_z)=(87/71,5,\sqrt{2} \pi/5).$ 
The corresponding numerical data are collected in tables \ref{text11}-\ref{text15}.

\begin{table}[h]
\begin{center}
\begin{tabular}{|c|c|c|}
\hline
$n$ & $ c_n^{(-1,\al)}$   & $ c_n^{(-3,\al)} $  \tabularnewline
 \hline
 $1$  & $0.047836895895342651045197466476518$ & $-0.092720704254336809423417304015711$    \\
 \hline
$2$  & $0.22271552228325462076432403940644$ & $-0.21799328810557764590768982278561$    \\
 \hline
$3$  & $0.33632324493557158980998094287678$ & $-0.092728756209299748309817148928123$    \\
 \hline
$4$  & $0.30443130992545063377032982859603$ & $0.030697006419762539926301146611732$    \\
 \hline
\end{tabular}\label{c13}
\bigskip
\caption{Numerical data for $c_n^{(-(2j-1),\al)}$ with $j=1,2$ and $n=1,2,3,4$, 
corresponding to the parameter choice $(\al,p,\al_z)=(87/71,5,\sqrt{2} \pi/5).$
}
\label{text11}
\end{center}
\end{table}
\normalsize

\begin{table}[h]
\begin{center}
\begin{tabular}{|c|c|c|}
\hline
$n$ & $ c_n^{(-5,\al)}$   & $ c_n^{(-7,\al)} $  \tabularnewline
 \hline
 $1$  & $1.0685900883396947464605339869010$ & $-30.244034919904560095845721663325$    \\
 \hline
$2$  & $2.0071148246050986412162379727146$ & $-48.944299373241089726706490008931$    \\
 \hline
$3$  & $0.64561760254807434224661704391942$ & $-12.973259214820955883108435935957$    \\
 \hline
$4$  & $-0.13724577423038151570814864027627$ & $1.7992348409013577383655892972328$    \\
 \hline
\end{tabular}\label{c57}
\bigskip
\caption{Numerical data for $c_n^{(-(2j-1),\al)}$ with $j=3,4$ and $n=1,2,3,4$, 
corresponding to the parameter choice $(\al,p,\al_z)=(87/71,5,\sqrt{2} \pi/5).$
}
\label{text12}
\end{center}
\end{table}
\normalsize

\begin{table}[h]
\begin{center}
\begin{tabular}{|c|c|c|}
\hline
$n$ & $ I_{n}^{(1)}$   & $ I_n^{(2)} $  \tabularnewline
 \hline
 $1$  & $-0.72291137954811674403064603025472$ & $1.4011956873585816721199852641882$   \\
 \hline
$2$  & $0.062026088692890819205809166968978$ & $-0.14893106352715074437275536320374$    \\
 \hline
$3$  & $3.2967971369860991667474730091582$ & $-1.3228398983910281935603450850790$   \\
 \hline
$4$  & $-9.1081390303390309923144091253807$ & $-1.9739047914098149774389138884225$    \\
 \hline
\end{tabular}\label{In12}
\bigskip
\caption{Numerical data for $I_n^{(k)}$ with $k=1,2$ and $n=1,2,3,4$, corresponding to the parameter choice $(\al,p,\al_z)=(87/71,5,\sqrt{2} \pi/5).$
}
\label{text13}
\end{center}
\end{table}
\normalsize

\begin{table}[h]
\begin{center}
\begin{tabular}{|c|c|c|}
\hline
$n$ & $ I_{n}^{(3)}$   & $ I_n^{(4)} $  \tabularnewline
 \hline
 $1$  & $-16.148538078707194010668860120792$ & $457.04798770562063166307628541906$    \\
 \hline
$2$  & $1.7842782303175010585218716255752$ & $-51.556266303265126299457434554926$   \\
 \hline
$3$  & $10.826245871586022756179909139045$ & $-248.44154167101785943203439381949$    \\
 \hline
$4$  & $12.900068157937816163666201714057$ & $-260.42758116035009175688379626119$    \\
 \hline
\end{tabular}\label{In34}
\bigskip
\caption{Numerical data for $I_n^{(k)}$ with $k=3,4$ and $n=1,2,3,4$ at the parameter values $(\al,p,\al_z)=(87/71,5,\sqrt{2} \pi/5).$
}
\label{text14}
\end{center}
\end{table}
\normalsize


\begin{table}[h]
\begin{center}
\begin{tabular}{|c|c|c|}
\hline
$m$ & $ {\cal C}(\al,m,\nu)\big|_{CFT}$   & $ {\cal C}(\al,m,\nu)\big|_{kinks}\!\!\!\!\!\!\!\!-{\cal C}(\al,m,\nu)\big|_{CFT}    $  \tabularnewline
 \hline
 $1$  & $-1.9415623381592084252901351919255$ & $2.7129047 \cdot 10^{-29}$   \\
 \hline
$2$  & $-0.37708224526339741051552387325847$ & $1.1717275 \cdot 10^{-28}$  
 \\
 \hline
$3$  & $28.789271296522235052927504568519$ & $1.5250733\cdot 10^{-25}$   
 \\
 \hline
$4$  & $131931.29437447900816353730073806$ & $-2.2392240 \cdot 10^{-19}$  
 \\
 \hline
\end{tabular}\label{omnum1}
\bigskip
\caption{Numerical values of $ {\cal C}(\al,m,\nu)$ computed from the analytical CFT formula (\ref{simVEV1}), and the differences with respect to the corresponding values obtained from the integrable formulation through numerical evaluation of (\ref{calCres}) at $(\al,p,\al_z)=(87/71,5,\sqrt{2}\pi/5)$, for $m=1,2,3,4$.
}
\label{text15}
\end{center}
\end{table}
\normalsize


Next, we chose the parameter values $(\al,p,\al_z)=(0,5.1,\sqrt{2} \pi/10).$ 
The corresponding numerical data are collected in tables \ref{text21}-\ref{text25}.

\begin{table}[h]
\begin{center}
\begin{tabular}{|c|c|c|}
\hline
$n$ & $ c_n^{(-1,\al)}$   & $ c_n^{(-3,\al)} $  \tabularnewline
 \hline
 $1$  & $0.11069492959882507218035715198630$ & $-0.172398906101475954700690869124987$    \\
 \hline
$2$  & $0.26630538090754222588623176094278$ & $-0.178538508112979177463049169402614$    \\
 \hline
$3$  & $0.32854152932417626947511314060951$ & $-0.008260335391355905018526471915161$    \\
 \hline
$4$  & $0.25548452475294477531664312415741$ & $0.082905676514686635679263555047399$    \\
 \hline
\end{tabular}\label{c132}
\bigskip
\caption{Numerical data for $c_n^{(-(2j-1),\al)}$ with $j=1,2$ and $n=1,2,3,4$, 
corresponding to the parameter choice $(\al,p,\al_z)=(0,5.1,\sqrt{2} \pi/10).$
}
\label{text21}
\end{center}
\end{table}
\normalsize

\begin{table}[h]
\begin{center}
\begin{tabular}{|c|c|c|}
\hline
$n$ & $ c_n^{(-5,\al)}$   & $ c_n^{(-7,\al)} $  \tabularnewline
 \hline
 $1$  & $1.827459448820906568163597167831592$ & $-48.101928030460522126914229204213$    \\
 \hline
$2$  & $1.517246006094595877323298410567924$ & $-34.411463375076030662047302466111$    \\
 \hline
$3$  & $0.054112782057451919326951852457920$ & $-1.0238436718797062066460200965936$    \\
 \hline
$4$  & $-0.387022453970091192776624143093103$ & $5.50640816737967584891779932840497$    \\
 \hline
\end{tabular}\label{c572}
\bigskip
\caption{Numerical data for $c_n^{(-(2j-1),\al)}$ with $j=3,4$ and $n=1,2,3,4$, 
corresponding to the parameter choice $(\al,p,\al_z)=(0,5.1,\sqrt{2} \pi/10).$
}
\label{text22}
\end{center}
\end{table}
\normalsize

\begin{table}[h]
\begin{center}
\begin{tabular}{|c|c|c|}
\hline
$n$ & $ I_{n}^{(1)}$   & $ I_n^{(2)} $  \tabularnewline
 \hline
 $1$  & $0.228436947573430804176600812490566$ & $-0.355773114609196069089935063558978$   \\
 \hline
$2$  & $1.904663462822469351846017192589260$ & $-1.374347002610423779900969643738635$    \\
 \hline
$3$  & $60.25430858517157507366578813772764$ & $-1.807722663254031899822398836166532$   \\
 \hline
$4$  & $-3.782023590665786861969887717491056$ & $-1.794778907795742534860179542903450$    \\
 \hline
\end{tabular}\label{In122}
\bigskip
\caption{Numerical data for $I_n^{(k)}$ with $k=1,2$ and $n=1,2,3,4$, corresponding to the parameter choice $(\al,p,\al_z)=(0,5.1,\sqrt{2} \pi/10).$
}
\label{text23}
\end{center}
\end{table}
\normalsize

\begin{table}[h]
\begin{center}
\begin{tabular}{|c|c|c|}
\hline
$n$ & $ I_{n}^{(3)}$   & $ I_n^{(4)} $  \tabularnewline
 \hline
 $1$  & $3.7712590794881642415228654272578$ & $-99.266133069494110712564235134682$    \\
 \hline
$2$  & $12.038915872614486598378336346355$ & $-279.64720074571123877297521974733$   \\
 \hline
$3$  & $12.753440735272802944901768405309$ & $-258.11374960269750277674847155242$    \\
 \hline
$4$  & $9.9675529399537060797716115285337$ & $-174.16271092899449140154287459559$    \\
 \hline
\end{tabular}\label{In342}
\bigskip
\caption{Numerical data for $I_n^{(k)}$ with $k=3,4$ and $n=1,2,3,4$ at the parameter values $(\al,p,\al_z)=(0,5.1,\sqrt{2} \pi/10).$
}
\label{text24}
\end{center}
\end{table}
\normalsize

\begin{table}[h]
\begin{center}
\begin{tabular}{|c|c|c|}
\hline
$m$ & $ {\cal C}(\al,m,\nu)\big|_{CFT}$   & $ {\cal C}(\al,m,\nu)\big|_{kinks}\!\!\!\!\!\!\!\!-{\cal C}(\al,m,\nu)\big|_{CFT}    $  \tabularnewline
 \hline
 $1$  & $-0.22329082545498075990783605406139$ & $-1.0632873 \cdot 10^{-46}$   \\
 \hline
$2$  & $2.23146223064373204498149816206575$ & $-4 \cdot 10^{-45}$  
 \\
 \hline
$3$  & $-6043.643717004401926217414325156247$ & $ \sim 10^{-32}$   
 \\
 \hline
$4$  & $3033688.49809616723778888072945165$ & $  \sim 10^{-13}$  
 \\
 \hline
\end{tabular}\label{omnum2}
\bigskip
\caption{Numerical values of $ {\cal C}(\al,m,\nu)$ computed from the analytical CFT formula (\ref{simVEV1}), and the differences with respect to the corresponding values obtained from the integrable formulation through numerical evaluation of (\ref{calCres}) at $(\al,p,\al_z)=(0,5.1,\sqrt{2}\pi/10)$, for $m=1,2,3,4$.
}
\label{text25}
\end{center}
\end{table}
\normalsize

We also checked the validity of our results in the attractive regime with the parameter values  $(\al,p,\al_z)=(0,0.72,\sqrt{2} \pi/10).$ 
The corresponding numerical data are collected in tables \ref{text31}-\ref{text35}.

\begin{table}[h]
\begin{center}
\begin{tabular}{|c|c|c|}
\hline
$n$ & $ c_n^{(-1,\al)}$   & $ c_n^{(-3,\al)} $  \tabularnewline
 \hline
 $1$  & $1.06418761493916026115362708579848$ & $0.213234390349378727651106151920039$    \\
 \hline
$2$  & $-0.11581269798366615865964995335036$ & $0.417071343474958846121577394997053$    \\
 \hline
$3$  & $0.02669412007169057044836159127859$ & $0.629034827055661889356171137682552$    \\
 \hline
$4$  & $-0.00514501176770031369064819500032$ & $-0.074455640522790742128162001138894$    \\
 \hline
\end{tabular}\label{c133}
\bigskip
\caption{Numerical data for $c_n^{(-(2j-1),\al)}$ with $j=1,2$ and $n=1,2,3,4$, 
corresponding to the parameter choice $(\al,p,\al_z)=(0,0.72,\sqrt{2} \pi/10).$
}
\label{text31}
\end{center}
\end{table}
\normalsize

\begin{table}[h]
\begin{center}
\begin{tabular}{|c|c|c|}
\hline
$n$ & $ c_n^{(-5,\al)}$   & $ c_n^{(-7,\al)} $  \tabularnewline
 \hline
 $1$  & $-1.356435392613837089538077729871199$ & $30.8253720349259393959093773512897$    \\
 \hline
$2$  & $-0.600715498215618378265777677987605$ & $3.87454546587688103724218940054885$    \\
 \hline
$3$  & $0.164863056813399585710020071216592$ & $0.15045638119735585338384309552279$    \\
 \hline
$4$  & $0.971029336115185556874873707053653$ & $-0.8249219522334881387410835953159$    \\
 \hline
\end{tabular}\label{c573}
\bigskip
\caption{Numerical data for $c_n^{(-(2j-1),\al)}$ with $j=3,4$ and $n=1,2,3,4$, 
corresponding to the parameter choice $(\al,p,\al_z)=(0,0.72,\sqrt{2} \pi/10).$
}
\label{text32}
\end{center}
\end{table}
\normalsize

\begin{table}[h]
\begin{center}
\begin{tabular}{|c|c|c|}
\hline
$n$ & $ I_{n}^{(1)}$   & $ I_n^{(2)} $  \tabularnewline
 \hline
 $1$  & $-6.482572562543416780942990308462397$ & $-1.298932057528767581801601716739920$   \\
 \hline
$2$  & $-0.646002370234018297756757981739038$ & $0.5643231340627016379902707986705130$    \\
 \hline
$3$  & $-0.180549396235322585646887215632594$ & $-2.170352302407947643890687909540439$   \\
 \hline
$4$  & $-0.054058145279772132054554061655150$ & $-0.435661918158286591442804026174149$    \\
 \hline
\end{tabular}\label{In123}
\bigskip
\caption{Numerical data for $I_n^{(k)}$ with $k=1,2$ and $n=1,2,3,4$, corresponding to the parameter choice $(\al,p,\al_z)=(0,0.72,\sqrt{2} \pi/10).$
}
\label{text33}
\end{center}
\end{table}
\normalsize

\begin{table}[h]
\begin{center}
\begin{tabular}{|c|c|c|}
\hline
$n$ & $ I_{n}^{(3)}$   & $ I_n^{(4)} $  \tabularnewline
 \hline
 $1$  & $8.2628201414691105499572679141060$ & $-187.77488873071333391233823449625$    \\
 \hline
$2$  & $-0.3557738989851967029431600665925$ & $-7.3208140161521685950478081973356$   \\
 \hline
$3$  & $-1.0546741160622623817622550128773$ & $2.11513829317075240051141579548565$    \\
 \hline
$4$  & $2.22431142187919984193921334941271$ & $-2.15048302127616256461928868093482$    \\
 \hline
\end{tabular}\label{In343}
\bigskip
\caption{Numerical data for $I_n^{(k)}$ with $k=3,4$ and $n=1,2,3,4$ at the parameter values $(\al,p,\al_z)=(0,0.72,\sqrt{2} \pi/10).$
}
\label{text34}
\end{center}
\end{table}
\normalsize

\begin{table}[h]
\begin{center}
\begin{tabular}{|c|c|c|}
\hline
$m$ & $ {\cal C}(\al,m,\nu)\big|_{CFT}$   & $ {\cal C}(\al,m,\nu)\big|_{kinks}\!\!\!\!\!\!\!\!-{\cal C}(\al,m,\nu)\big|_{CFT}    $  \tabularnewline
 \hline
 $1$  & $2.52148185536943272168970068805008$ & $2.1942354 \cdot 10^{-62}$   \\
 \hline
$2$  & $0.32230778564177332484258690140558$ & $-1.0138852 \cdot 10^{-58}$  
 \\
 \hline
$3$  & $0.58262538308093700850639042863797$ & $-8.6208462\cdot 10^{-58}$   
 \\
 \hline
$4$  & $-4.711348085331561347687751423160431$ & $ \sim 10^{-40}$  
 \\
 \hline
\end{tabular}\label{omnum13}
\bigskip
\caption{Numerical values of $ {\cal C}(\al,m,\nu)$ computed from the analytical CFT formula (\ref{simVEV1}), and the differences with respect to the corresponding values obtained from the integrable formulation through numerical evaluation of (\ref{calCres}) at $(\al,p,\al_z)=(0,0.72,\sqrt{2}\pi/10)$, for $m=1,2,3,4$.
}
\label{text35}
\end{center}
\end{table}
\normalsize



In each table, in order to keep the size of the presentation manageable, we display only approximately 33 digits of the numerical values, even in cases where our numerical method yields substantially higher precision.

\newpage

\clearpage





\newpage

\providecommand{\href}[2]{#2}\begingroup\raggedright\endgroup

\end{document}